\documentclass[final]{siamart171218}

\usepackage{colortbl}
\usepackage{graphics,graphicx,color,arydshln,amsfonts,amsopn,braket,subfig}
\usepackage{subfig}
\usepackage{bbm}

\newcommand{\mysolutionb}{\begin{solution} \color{blue}}
\newcommand{\mysolutione}{\end{solution}}

\newcommand{\btbl}{\begin{tabular}{l}}
\newcommand{\et}{\end{tabular}}

\newcommand{\req}[1]{\cref{#1.eq}}
\newcommand{\be}{\begin{equation}}
\newcommand{\ee}{\end{equation}}
\newcommand{\Z}{{\mathbb Z}}
\newcommand{\R}{{\mathbb R}}
\newcommand{\C}{{\mathbb C}}
\newcommand{\T}{{\mathbb T}}
\newcommand{\G}{{\mathbb G}}
\newcommand{\N}{{\mathbb N}}
\newcommand{\mbS}{{\mathbb S}}

\newcommand{\cM}{{\cal M}}

        \newcommand{\matbegin}{   	\left[		}
        \newcommand{\matend}{	\right]	}

	\newcommand{\matparbegin}{
        \renewcommand{\baselinestretch}{1}
        \renewcommand{\arraystretch}{.5}
        \setlength{\arraycolsep}{.25em}
        \left[
        }
        \newcommand{\matparend}{
            \right]
        }

	\newcommand{\obthdtall}[3]{
 		\matparbegin \begin{array}{c:c:c}
       	 	 \rule{0em}{1em}  &   & \rule{0em}{0em}  \\  
               	#1 & #2 & #3  \\  
        		  \rule{0em}{1em}  & & \rule{0em}{0em}
		\end{array}\matparend}
        \newcommand{\thbthd}[9]{
         	\matbegin \begin{array}{c:c:c}
                	#1 & #2 & #3     {\rule{0em}{.1em}}_{\rule{0em}{.7em}}   \\  \hdashline
                        #4 & #5 & #6 \\  \hdashline
                	#7 & #8 & #9 \rule{0em}{1em}
        	\end{array}\matend}
		
\newcommand{\bbm}{\begin{bmatrix}}
\newcommand{\ebm}{\end{bmatrix}}
\newcommand{\bsm}{\left[ \begin{smallmatrix}}
\newcommand{\esm}{ \end{smallmatrix} \right]}
\newcommand{\lb}{\left(}
\newcommand{\rb}{\right)}
\newcommand{\lcb}{\left\{}
\newcommand{\rcb}{\right\}}
\newcommand{\xh}{\hat{x}}

\newcommand{\ah}{\hat{a}}

\newcommand{\sm}{\text{-}}

\newcommand{\colb}{\color{blue}}
\newcommand{\tcb}[1]{\textcolor{blue}{#1}}

\newcommand{\diago}[1]{{\rm diag}\left( #1 \right)} 	
\newcommand{\diag}[1]{{\rm diag}\left( #1 \right)} 	

\newcommand{\cV}{{\mathcal V}}
\newcommand{\cH}{{\mathcal H}}

\newcommand{\GL}{{\rm GL}}

\newcommand{\otbm}[9]{	
		\left[ \begin{array}{c:cc}
		#1 & #2 & #3 \\ \hdashline #4 & #5 & #6 \\ #7 & #8 & #9 \end{array} \right] }
\newcommand{\otbms}[9]{	\footnotesize		\arraycolsep=1.8pt\def\arraystretch{1}
		\left[ \begin{array}{c:cc}
		#1 & #2 & #3 \\ \hdashline #4 & #5 & #6 \\ #7 & #8 & #9 \end{array} \right] }

\newcommand{\tbtdash}[4]{			
		\left[ \begin{array}{c:c}
		#1 & #2 \\ \hdashline #3 & #4  \end{array} \right] }

\newcommand{\inprod}[2]{\left< #1 , #2 \right>}

\newcommand{\mycaption}[1]{\caption{\footnotesize #1}}

\title{Discovering Transforms: \\ A Tutorial on  Circulant Matrices,  Circular Convolution, and the Discrete Fourier Transform} 

\author{Bassam Bamieh\thanks{Department of Mechanical Engineering, University of California at Santa Barbara, {\em bamieh@ucsb.edu.} This work is partially supported by NSF Awards CMMI-1763064 and ECCS-1932777.}}

\begin{document}

\maketitle 



\begin{abstract} 
	How could the Fourier and other transforms be naturally discovered  if one didn't know how to postulate them? In the case of the Discrete Fourier Transform (DFT), we show how it arises naturally out of analysis of circulant matrices. In particular, the DFT can be derived as the change of basis that simultaneously diagonalizes all circulant matrices.  In this way,   the DFT arises naturally from a linear algebra question about a set of matrices. Rather than thinking of the DFT as a signal transform, it is  more natural to think of it as a single change of basis that renders an entire  set of mutually-commuting matrices into simple, diagonal forms. The DFT can then be ``discovered'' by solving the eigenvalue/eigenvector problem for a special element in that set. A  brief outline is given  of how this line of thinking can be generalized to families of linear operators, leading to the discovery of  the other  common Fourier-type transforms. 	
\end{abstract}

%
%


\section{Introduction} 

The Fourier transform in all its forms is ubiquitous. Its many useful properties are introduced early on in  Mathematics, Science and Engineering curricula~\cite{oppenheim1983signals}. Typically, it is introduced as a transformation on functions or signals, and then its many useful  properties are easily derived. Those properties are then shown to be remarkably effective in solving certain differential equations, or in analyzing the action of time-invariant linear dynamical systems, amongst many other uses.   To the student, the effectiveness of the Fourier transform in solving these problems may seem magical at first, before familiarity eventually suppresses that initial sense of wonder. In this tutorial, I'd like to step back to before one is shown the Fourier transform, and ask the following question:  How would one naturally {\em discover} the Fourier transform rather than have it be {\em postulated}? 

The above question is interesting for several reasons. First, it is more intellectually satisfying to introduce a new mathematical object from familiar and well-known objects rather than having it postulated ``out of thin air''. In this tutorial we demonstrate how the DFT arises naturally from the problem of {\em simultaneous diagonalization} of all circulant matrices, which share symmetry properties that enable this diagonalization. It should be noted that simultaneous diagonalization of any class of linear operators or matrices is the ultimate way to understand their actions, by reducing the entire class to the simplest form of linear operations (diagonal matrices) simultaneously. The same procedure can be applied to discover the other close relatives of the DFT, namely the Fourier Transform, the $z$-Transform and Fourier Series. All can be arrived at by simultaneously diagonalizing a respective class of linear operators that obey their respective symmetry rules. 

To make the point above, and to have a concrete discussion,  in this tutorial we consider primarily the case of circulant matrices. This case is also particularly useful because it yields the DFT, which is the computational workhorse for all Fourier-type analysis. 
Given an  $n$-vector $a~:=~(a_0,\ldots,a_{n-1})$, 
define the associated  matrix $C_a$ whose 
first column is made up of these numbers, and each subsequent column is obtained  by a {\em circular shift} of the previous column
\be
	C_a ~:=~\bbm
			a_0 & a_{n\sm 1} & a_{n\sm 2} & \cdots & a_{1}	\\
			a_{1} & a_0 & a_{n\sm 1} &              & a_{2}	\\
			a_{2} & a_{1} & a_0 &              & a_{3}	\\
			\vdots     &  &\ddots &   \ddots           & \vdots	\\
			a_{n\sm 1}    & a_{n\sm 2} & a_{n\sm  3} &   \cdots           & a_0
		\ebm .
   \label{circ_def.eq}
\ee
Note that each row is also obtained from the pervious row by a circular shift. Thus the entire matrix is completely determined by any one of its rows or columns. 
Such matrices are called {\em circulant}. They are a subclass of Toeplitz matrices, and as mentioned,  have  very special properties 
due to their intimate relation to the Discrete Fourier Transform (DFT) and circular convolution. 

Given an $n$-vector $a$ as above, its DFT $\ah$ is another $n$-vector defined by 
\be
	\ah_k ~:=~ \sum_{l=0}^{n-1} a_l ~e^{-i\frac{2\pi}{n} kl } , ~~~~~~k=0,1,\ldots, n- 1. 
  \label{DFT_def.eq}	
\ee
A remarkable fact is that given a circulant matrix $C_a$, its eigenvalues are easily computed. They are precisely the set of complex numbers $\lcb \ah_k\rcb$, i.e. the DFT of the vector $a$ that defines the circulant matrix $C_a$. There are many ways to derive this conclusion and other  properties of the DFT. Most treatments start with the definition~\req{DFT_def} of the DFT, from which many of its seemingly magical properties are easily derived. To restate the goal of this tutorial, the question we ask here is: what if we didn't know  the DFT? How can we arrive at it in a natural manner without needing someone to postulate~\req{DFT_def} for us? 

There is a  natural way to think about this problem. Given a class of matrices or operators, one asks if there is a  transformation, a change of basis, in which their matrix representations  all have the same structure such as diagonal, block diagonal, or other special forms. The simplest such scenario is when a class of matrices can be {\em simultaneously diagonalized} with the same transformation. Since diagonalizing transformations are made up of eigenvectors of a matrix, then {\em a set of matrices is simultaneously diagonalizable iff they share a full set of eigenvectors}. An equivalent condition is that they each are diagonalizable, and they all mutually commute. Therefore given a mutually commuting set of matrices, by finding their shared eigenvectors, one finds that special transformation that simultaneously diagonalizes all of them. Thus, finding the ``right transform'' for a particular class of operators amounts to identifying the correct eigenvalue problem, and then calculating the eigenvectors, which then yield the transform. 

An alternative but complementary view of the above procedure involves describing the class of operators using some underlying common symmetry. For example,  circulant matrices such as~\req{circ_def}  have a shift invariance property with respect to circular shifts of vectors. This can also be described as having a shift-invariant action on vectors over $\Z_n$ (the integers modulo $n$), which is also equivalent to having a shift-invariant action on periodic functions (with period $n$). In more formal language, circulant matrices represent a class of mutually commuting operators that also commute with the action of the group $\Z_n$. A basic shift operator generates that group, and the eigenvalue problem for that shift operator yields the DFT. This approach has the advantage of being generalizable to more complex symmetries that can be encoded in the action of other, possibly non-commutative, groups. These techniques are part of the theory of group representations. However, we adopt here the approach described in the previous paragraph, which uses familiar Linear Algebra language and avoids the formalism of group representations. None the less, the two approaches are intimately linked. Perhaps the present approach can be thought of as a ``gateway'' treatment on a slippery slope to  group representations~\cite{plymen2010noncommutative,taylor1986noncommutative} if the reader is so inclined.


This tutorial follows the ideas described earlier. We first (\cref{diag.sec}) investigate the simultaneous diagonalization problem for matrices, which is of interest in itself, and show how it can be done constructively.  We then (\cref{circ_struct.sec}) introduce circulant matrices, explore their underlying geometric and symmetry properties, as well as their simple correspondence with circular convolutions. The general procedure for commuting matrices is then used (\cref{sdiag.sec}) for the particular case of circulant matrices to simultaneously diagonalize them. The traditionally defined DFT emerges naturally out of this procedure, as well as other equivalent transforms.  The ``big picture'' for the DFT is then summarized (\cref{big.sec}). A much larger context is briefly outlined in~\cref{further.sec}, where the close relatives of the DFT, namely the Fourier transform, the $z$-transform and Fourier series are discussed. Those can be arrived at naturally by simultaneously ``diagonalizing'' families of mutually commuting linear operators. In this case, diagonalization has to be interpreted in a more general sense of conversion to so-called multiplication operators. Finally (\cref{s3.sec}), an example of a non-commutative case is given where not diagonalization, but rather simultaneous {\em  block-diagonalization} is possible. This  serves as a motivation for generalizing  classical Fourier analysis to  so-called non-commutative Fourier analysis which is very much the subject of group representations.  

\section{Simultaneous Diagonalization of Commuting Matrices} 					\label{diag.sec}

The simplest matrices to study and understand are the diagonal matrices. They are basically  uncoupled sets of scalar multiplications, essentially the simplest of all possible linear operations. When a matrix $M$ can be diagonalized with a similarity transformation (i.e. $\Lambda =V^{-1} M V$, where $\Lambda$ is diagonal), then we have a change of basis in which the linear transformation has that simple diagonal matrix representation, and its properties can be easily understood. 

Often one has to work with a {\em set of transformations} rather than a single one, and usually with sums and products of elements of that set. If we require a different similarity transformation for each member of that set, then sums and products will each require finding their own diagonalizing transformation, which is a lot of work.  It is then natural to ask if there exists one basis in which all members of a set of transformations have diagonal forms. This is the simultaneous diagonalization problem.  If such a basis exists, then the properties of the entire set, as well as all sums and products (i.e. the algebra generated by that set) can be easily deduced from their diagonal forms. 
\begin{definition} 
	A set $\cM$ of matrices is called {\em simultaneously diagonalizable} if there exists a single similarity transformation that diagonalizes all matrices in $\cM$. In other words, there exists a single non-singular matrix $V$, such that for each $M\in\cM$, the matrix 
	\[
		V^{-1}MV ~=~ \Lambda ~~~~~\mbox{is diagonal}. 
	\]
\end{definition}
It is immediate that all sums, products and inverses (when they exist) of elements of $\cM$ will then  also be diagonalized by this same similarity transformation. Thus a simultaneously diagonalizing transformation, when it exists,  would be an invaluable tool in studying such sets of matrices. 

When can a given set of matrices be simultaneously diagonalized? The answer is  simple to state. First, they each have to be individually diagonalizable as an obvious necessary condition. Then, we will show that {\em a set of diagonalizable matrices can be simultaneously diagonalized iff they all mutually commute}. We will illustrate the argument in some detail since it gives a procedure for constructing the diagonalizing transformation. In the case of circulant matrices, this construction will yield the DFT. We note that the same construction also yields the z-transform, Fourier transform and Fourier series, but with some slight additional technicalities due to working with operators on infinite-dimensional spaces. 

\textit{Necessity:}  We can see that commutativity is a necessary condition because all diagonal matrices mutually commute, and if two matrices are simultaneously diagonalizable, they do commute in the new basis, and therefore they must commute in the original basis. More precisely, let $A=V^{-1}\Lambda_a V$ and $B=V^{-1}\Lambda_b V$ be simultaneously diagonalizable with the transformation $V$, then 
\begin{align} 
	AB &= \big( V^{-1} \Lambda_a V\big) \big(  V^{-1} \Lambda_b V \big)
			= V^{-1} \Lambda_a  \Lambda_b V
			= V^{-1}  \Lambda_b \Lambda_a V									\nonumber		\\
			&= \big( V^{-1}  \Lambda_b V \big) \big( V^{-1}  \Lambda_a V \big) 	 
			= BA														\label{sim_com.eq}
\end{align}

What about the converse? If two matrices commute, are they simultaneously diagonalizable? The answer is yes if both matrices 
are diagonalizable individually (this is of course a necessary condition). The argument is simple if one of the matrices has non-repeated eigenvalues. A little more care needs to be taken in the case of repeated eigenvalues since there are many diagonalizing transformations in that case.  We will not need the more general version of this argument in this tutorial. 

To begin, let's recap how one constructively diagonalizes a given matrix by finding its eigenvectors. If $v_i$ is a an eigenvector of an $n\times n$ matrix $A$ with corresponding eigenvalue $\lambda_i$ then we have  
\be
	Av_i ~=~ \lambda_i v_i , ~~~~~i=1,\ldots, p, 
  \label{eigs.eq}
\ee
where $p$ is the largest number of linearly independent eigenvectors (which can be any number from $1$ to $n$). The relations~\req{eigs}
can be compactly rewritten using {\em partitioned matrix notation}  as a single matrix equation 
	\begin{align}
		 \obthdtall{Av_1}{\cdots}{Av_p}
		 &=  
		\obthdtall{\lambda_1 v_1}{\cdots}{\lambda_m v_p} 										\nonumber	\\
		& \Updownarrow 																	\nonumber	\\
		\bbm & & \\ & A & \\ & & \ebm \obthdtall{v_1}{\cdots}{v_p} 
		& =  
		 \obthdtall{v_1}{\cdots}{v_p}
		\thbthd{\lambda_1}{}{}{}{\ddots}{}{}{}{\lambda_p}		
		~~~\Leftrightarrow~~~			 	   				
		AV  =  V \Lambda, 																\label{AV.eq}	
	\end{align}
	where $V$ is a matrix whose columns are the eigenvectors of $A$, and $\Lambda$ 
	is the diagonal matrix made up of the corresponding eigenvalues of $A$.  
	
	We say that an $n\times n$ matrix has a {\em full set of eigenvectors} if it has $n$ linearly independent eigenvectors. 
	In that case, the matrix $V$ in~\req{AV} is square and nonsingular and 
	$
		\Lambda ~=~ V^{-1} A V 
	$
	is the diagonalizing similarity transformation. Of course not all matrices have a full set of eigenvectors. If the Jordan form 
	of a matrix contains any non-trivial Jordan blocks, then it can't be diagonalized, and has strictly less than $n$ linearly 
	independent eigenvectors. We can therefore state that {\em a matrix is diagonalizable iff it has a full set of eigenvectors}, 
	i.e. diagonalization is equivalent (in a constructive sense) to finding $n$ linearly independent eigenvectors. 
	
	\subsubsection*{The case of simple (non-repeated) eigenvalues} 
	
	Now consider the problem of simultaneous diagonalization. It is clear from the above discussion that two matrices
	can be simultaneously diagonalized iff they {\em share a full set of eigenvectors}. Consider the converse of the
	argument~\req{sim_com}, and assume  that $A$ has (simple) non-repeated eigenvalues. This means that 
	\[
		A v_i ~=~ \lambda_i v_i, ~~~~i=1,\ldots,n, ~~~\mbox{and } \lambda_i\neq\lambda_j \mbox{ if } i\neq j. 
	\]
	Consider any matrix  $B$ that commutes with $A$. Let $B$ act on each of the eigenvectors by $Bv_i$ and observe that 
	\be
		A ~\big( Bv_i \big) ~=~ B ~A v_i ~=~ B ~\lambda_i v_i ~=~ \lambda_i ~\big(B v_i \big) .
	   \label{distinct_arg.eq}
	\ee
	Thus $Bv_i$ is an eigenvector of $A$ with eigenvalue $\lambda_i$. Since those eigenvalues are distinct, and the
	corresponding eigenspace is one dimensional, $Bv_i$ must be a scalar multiple of $v_i$
	\[
		Bv_i ~=~ \gamma_i v_i. 
	\]
	Thus $v_i$ is an eigenvector of $B$, but possibly with an eigenvalue $\gamma_i$ different from $\lambda_i$.
	In other words, the eigenvectors of $B$ are exactly the unique (up to scalar multiples) eigenvectors of $A$. 
	We summarize this next. 
	\begin{lemma} 												\label{s_diag.lemma}
		If a matrix $A$ has simple eigenvalues, then $A$ and $B$ are simultaneously diagonalizable iff they commute.
		In that case, the diagonalizing basis is made up of the eigenvectors of $A$.
	\end{lemma} 
	This statement gives a constructive procedure 
	for simultaneously diagonalizing a set $\cM$ of mutually commuting matrices. If we can find one matrix  $A\in\cM$ 
	with simple eigenvalues, then find its eigenvectors, those will yield the simultaneously diagonalizing transformation 
	for the entire set. This is the procedure used for circulant matrices in~\cref{sdiag.sec}, where the ``shift operator'' 
	$S$ or its adjoint $S^*$ play the role of the matrix with simple eigenvalues. The diagonalizing transformation for
	$S^*$ yields the standard DFT. We will see that we can also produce other, equivalent versions of the DFT if we 
	use eigenvectors of $S$ instead, or eigenvectors of $S^p$ with $(p,n)$ co-prime.

\section{Structural Properties of Circulant Matrices} 						\label{circ_struct.sec}

The structure of circulant matrices is most clearly expressed using modular arithmetic. In some sense, modular arithmetic ``encods'' the symmetry  properties of circulant matrices. We begin with a geometric view of modular arithmetic  by relating it to rotations of roots of unity. We then show the ``rotation invariance'' of the action of circulant matrices, and finally connect that with circular convolution.

\subsection{Modular Arithmetic, $\Z_n$, and Circular Shifts}

To understand the symmetry properties of circulant matrices, it is useful to first study and  establish some simple properties of the set $\Z_n:=\lcb 0, 1, \cdots, n-1 \rcb$ of  integers {\em modulo $n$}. The arithmetic in $\Z_n$ is {\em modular arithmetic}, that is,  we say {\em $k$ equals $l$  modulo $n$} if   $k-l$ is an integer multiple of $n$. The following notation can be used to describe this formally 
\[
	\begin{array}{rll}  k &  \!\! = \!\! & l ~(\text{mod}~n)		\\
			\mbox{or~~}	k & \!\!  \equiv_n  \!\!  & l 		\end{array} 
	~~~~~~~~~\Longleftrightarrow~~~~~~~~~
	\exists m\in\Z, ~\text{s.t.}~~~~ k-l~=~ m ~n
\]
Thus for example $n \equiv_n 0$, and $n+1 \equiv_n 1$ and so on. There are two equivalent ways to define (and think) about $\Z_n$, one mathematically formal and the other graphical. The first is to consider the set of all integers $\Z$ and regard any two integers $k$ and $l$ such that $k-l$ is a multiple of $n$ as equivalent, or more precisely as members of the same {\em equivalence class}. The infinite set of integers $\Z$ becomes a finite set of equivalence classes with this equivalence relation.  
\begin{figure}[h]
		\centering
			\subfloat[]{\label{Z_n:bins.fig} \raisebox{1em}{\includegraphics[width=0.22\textwidth]{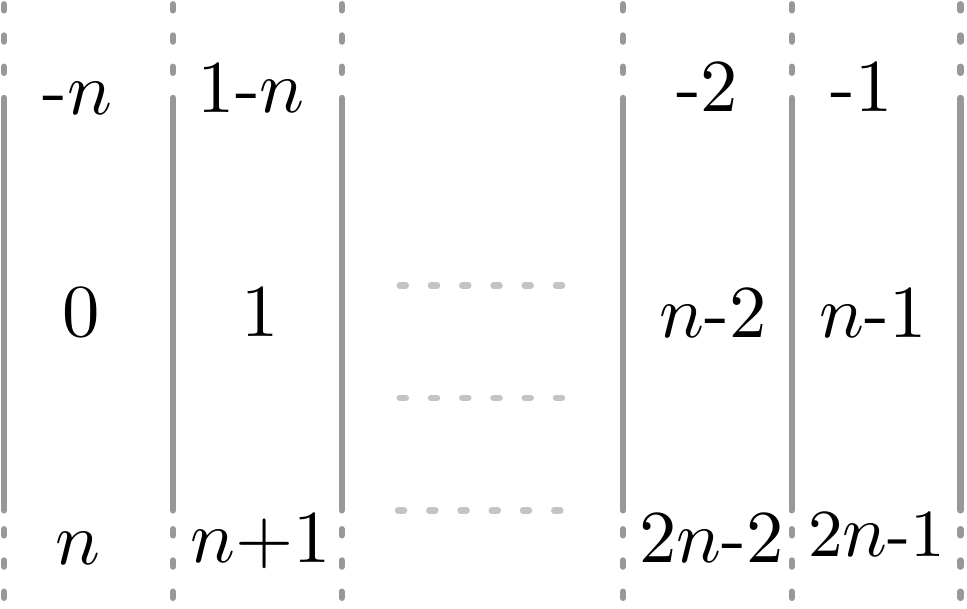}	}	}
			\hfill
			\subfloat[]{\label{Z_n:circle1.fig}	\includegraphics[width=0.22\textwidth]{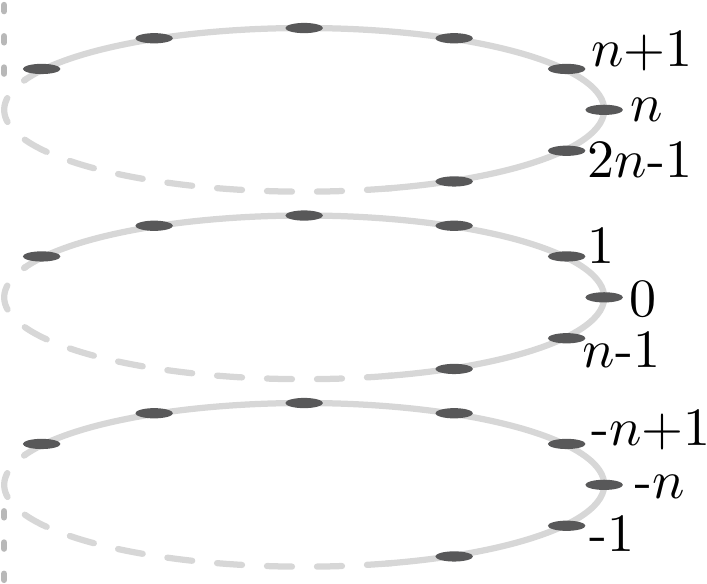}	}
			\hfill
			\subfloat[]{\label{Z_n:circle2.fig}	\raisebox{.5em}{\includegraphics[width=0.22\textwidth]{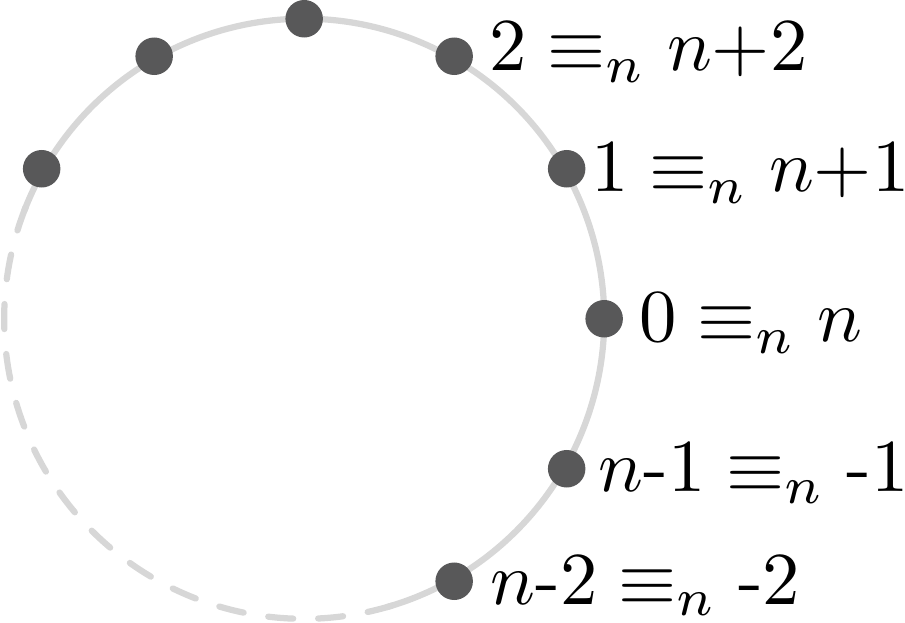}	}	}
			\hfill
			\subfloat[]{\label{Z_n:roots.fig}	\includegraphics[width=0.22\textwidth]{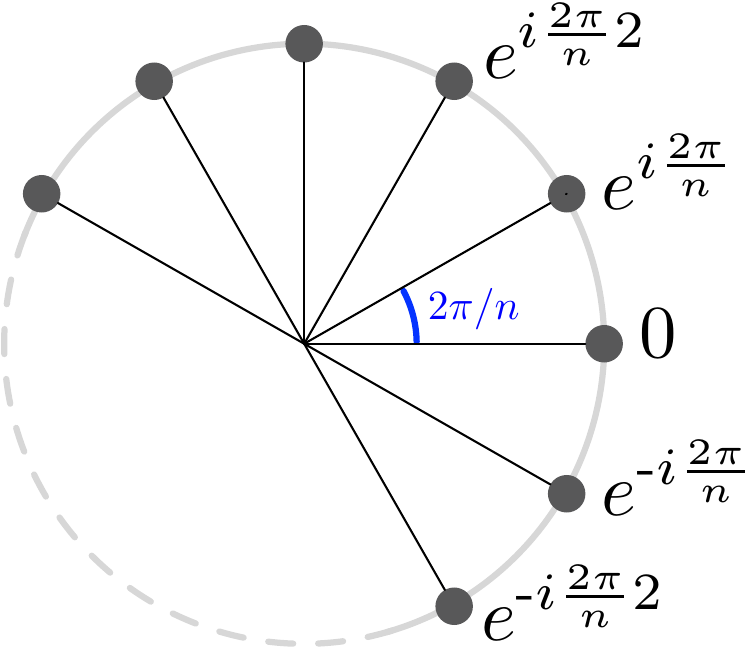}		}
	\mycaption{{ (a)} Definition of $\Z_n$ as the decomposition of the integers $\Z$ into  {\em equivalence classes} each indicated as a ``vertical bin''. Two integers in $\Z$ belong to the same equivalence class (and represent the same element of $\Z_n$) if they differ by an integer multiple of $n$. { (b)} Another depiction of the decomposition where two integers that are vertically aligned in this figure belong to the same equivalence class. The arithmetic in $\Z_n$ is just angle addition in this diagram. For example, $(-1)+(n+1) \equiv_n 0 \equiv_n n$. { (c)} Using the set $\lcb 0, 1, \cdots, n-1 \rcb$ as $\Z_n$. A few equivalent members are shown, and the arithmetic of $Z_n$ is just angle addition here. This can be thought of as the ``top view'' of (b). { (d)} The $n$th roots of unity $\rho_m:=\exp\lb {i\frac{2\pi}{n}m} \rb$ lying on the unit circle in the complex plane. Identifying $\rho_m$ with $m\in\Z_n$ shows that complex multiplication on $\lcb \rho_m\rcb$ (which corresponds to angle addition) is equivalent to modular addition in $\Z_n$.  }
	\label{Z_n.fig}
\end{figure}
This is illustrated in~\cref{Z_n:bins.fig} where elements of $\Z_n$ are arranged in ``vertical bins'' which are the equivalence classes. Each equivalence class can be identified with any of its members. One choice is to identify the first one with the element $0$, the second one with $1$, and so on up to the $n$'th class identified with the integer $n-1$. 
\cref{Z_n:circle2.fig} also shows how elements of $\Z_n$ can be arranged on a discrete circle so that the arithmetic in $\Z_n$ is identified with angle addition. One more useful isomorphism is between $\Z_n$ and the $n$th roots of unity $\rho_m:=e^{i\frac{2\pi}{n}m}$, $m=0,\ldots,n-1$. The complex numbers $\lcb \rho_m \rcb$ lie on the unit circle each at a corresponding angle of $\frac{2\pi}{n}m$ counter-clockwise from the real axis (\cref{Z_n:roots.fig}). Complex multiplication on $\lcb \rho_m \rcb$ corresponds to addition of their corresponding angles, and the mapping $\rho_m \rightarrow m$ is an isomorphism from complex multiplication on $\lcb \rho_m \rcb$ to modular arithmetic in $\Z_n$. 

Using modular arithmetic, we can write down the definition of a circulant matrix~\req{circ_def} by specifying the $kl$'th entry\footnote{Here, and in this entire tutorial, matrix rows and columns are indexed from $0$ to $n-1$ rather than the more traditional $1$ through $n$ indexing. This alternative indexing significantly simplifies notation, and corresponds more directly to modular arithmetic. 	} 
of the matrix $C_a$  as 
\be
	\lb C_a \rb_{kl} ~:=~ a_{k-l},  	~~~k,l\in\Z_n,
  \label{circ_def_ind.eq}	
\ee
where we use (mod $n$) arithmetic for computing $k-l$. It is clear that with this definition, the first column of $C_a$ is just the sequence $a_0,a_1,\cdots, a_{n-1}$. The second column is given by the sequence $\lcb a_{k-1}\rcb$ and is thus $a_{-1}, a_0, \cdots, a_{n-2}$, which is exactly the sequence $a_{n-1},a_0,\cdots,a_{n-2}$, i.e. a circular  shift of the first column. Similarly each subsequent column is a circular  shift of the column preceding it. 

\begin{figure}[h]
		\centering
		\subfloat[]{\includegraphics[height=0.08\textheight]{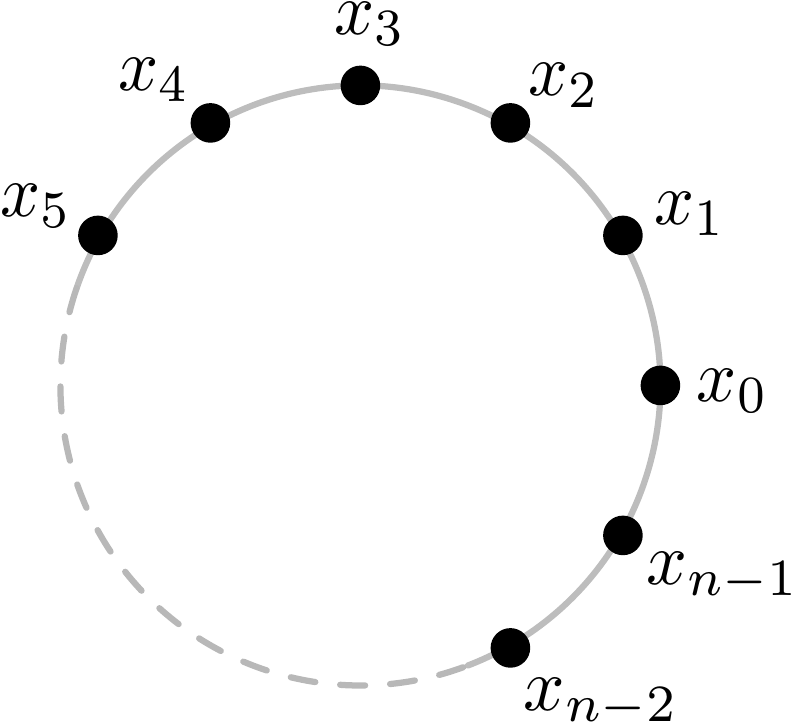}	}
		\hfill
		\subfloat[]{\includegraphics[height=0.08\textheight]{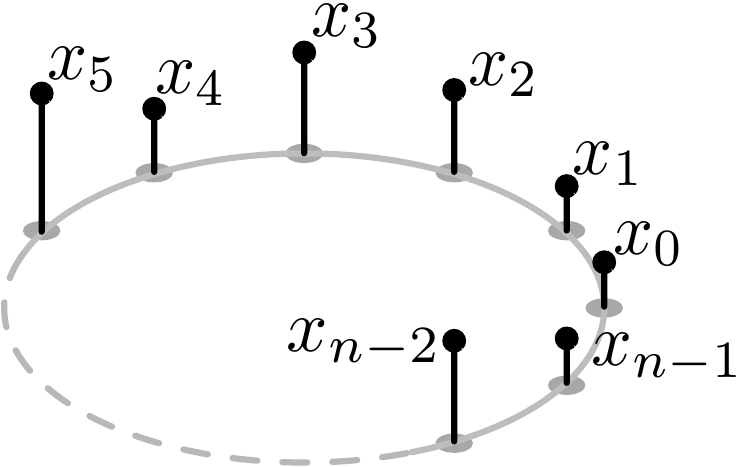}	}
		\hfill
		\subfloat[]{\raisebox{.5em}{	\includegraphics[height=0.06\textheight]{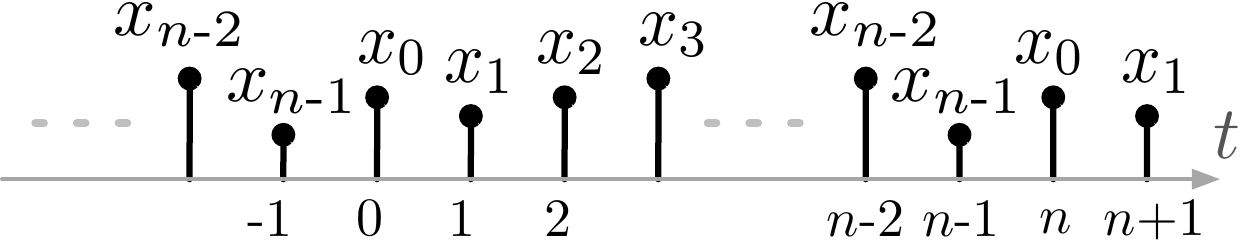}	}	}
	\mycaption{A vector $x :=(x_0,\ldots,x_{n-1})$ visualized as { (a)} a set of numbers arranged counter-clockwise  on a discrete circle, or equivalently { (b)} as a function $x:\Z_n\longrightarrow \C$ on the discrete circle $\Z_n$. (c) An $n$-periodic function on the integers $\Z$ can equivalently be viewed as a function on $\Z_n$ as in (b). }
	\label{circ_func.fig}
\end{figure}

Finally, it is useful to visualize an $n$-vector  $x:=(x_0,\ldots,x_{n-1})$ as a set of  numbers  
arranged at equidistant points along a circle, or equivalently as a {\em function} on the discrete circle. This is illustrated in~\cref{circ_func.fig}. 
Note the difference between this figure and~\cref{Z_n.fig}, which depicts the elements of $\Z_n$ and modular arithmetic. \cref{circ_func.fig} instead depicts {\em vectors} as a set of numbers arranged in a discrete circle, or as {\em functions on $\Z_n$}. A function on $\Z_n$ can also be thought of as a {\em periodic} function (with period $n$) on the set of integers $\Z$ (\cref{circ_func.fig}.c). In this case, periodicity of the function is expressed by the condition 
\be
	x_{t+n}=x_t ~~~\mbox{for all $t\in\Z$}. 
  \label{periodic.eq}
\ee
It is however more natural to view   periodic functions on $\Z$ as just functions on $\Z_n$. In this case, periodicity of the function is simply ``encoded'' in the modular arithmetic of $\Z_n$, and condition~\req{periodic} does not need to be explicitly stated.

\subsection{Symmetry Properties of Circulant Matrices}

Amongst all circulant matrices, there is a special one. 
Let $S$ and its adjoint $S^*$  be the {\em circular shift operators} defined by the following action on vectors  
\begin{align*}
	S~ \lb x_0  ,  ~\cdots , ~x_{n\sm 2} ,~ x_{n\sm 1}\rb 
				&=~ \lb x_{n\sm 1} ,~ x_0 ,~ \cdots ,~ x_{n\sm 2} \rb				\\
	S^*~\lb x_0 ,~ x_1  ,~ \cdots  ,~ x_{n\sm 1} \rb 
				&=~ \lb x_{1}  ,~   \cdots ,~ x_{n\sm 1},~ x_{0} \rb . 
\end{align*}
$S$ is therefore called the {\em circular right-shift operator} while $S^*$ is the {\em circular left-shift operator}. 
It is clear that $S^*$ is the inverse of $S$, and it is easy to show that it is the adjoint of $S$. The latter fact  also becomes clear upon examining the matrix representations of $S$ and $S^*$ 
\[	\small
	Sx = 
		\bbm 
			0 	& 		&  		& 	1	\\
			1	& 	&   	& 		\\
				&  	\ddots	&   \ddots	&  	 	\\
			    	& 	         &	1	&	 0		
		\ebm
	\bbm x_0 \\ x_1 \\  \vdots \\ x_{n\sm 1} \ebm		
	 =
	\bbm x_{n\sm 1} \\  x_0 \\  \vdots \\ x_{n\sm 2}  \ebm	, 
	~~
	S^*x = 
		\bbm 
			0 	& 1 		&  		& 		\\
				& \ddots	& \ddots  	& 		\\
				&  		&   	&  	 1	\\
			1    	& 	         &		&	 0		
		\ebm
	\bbm x_0 \\ x_1 \\  \vdots \\ x_{n\sm 1} \ebm		
	 =
	\bbm x_{1} \\  \vdots \\ x_{n\sm 1} \\  x_{0} \ebm	, 
\]	
which shows that $S^*$ is indeed the transpose (and therefore the adjoint) of $S$. Note that both matrix representations are circulant matrices since $S=C_{\lb 0,1,0,\ldots,0\rb}$ and  $S^*=C_{\lb 0,\ldots,0,1\rb}$ in the notation of~\req{circ_def}. The actions of $S$ and $S^*$    expressed  in terms of vector indices are
\be
	\lb Sx \rb_k ~:=~ x_{k-1} ,~~~~
	\lb S^*x \rb_k ~:=~ x_{k+1} , ~~~~~~~~~~~~k\in\Z_n,
  \label{S_ind.eq}
\ee
where modular  arithmetic is used for computing  vector indices.  For example $\lb Sx\rb_{0} =x_{0-1}  \equiv_n x_{n\sm 1}$. 

An important property of $S$ is that it commutes with any circulant matrix. One way to see this is to observe the for any matrix $M$, 
left (right) multiplication by $S$ amounts to row (column) circular permutation. A brief look at the circulant structure in~\req{circ_def}
shows that a row circular permutation gives the same matrix as a column circular permutation. Therefore, for any circulant matrix 
$C_a$, we have $SC_a=C_aS$. A more detailed argument is as follows.

To see this, note that the matrix representation of $S$ implies its $ij$'th entry is given by $\lb S\rb_{ij} = \delta_{i-j-1}$. Now let $C_a$ be any circulant matrix, and observe that 
\begin{align*}
	\lb  S C_a \rb_{ij}	& = \sum_l S_{il} \lb C_a \rb_{lj} = \sum_l \delta_{i-l-1} ~ a_{l-j} = \sum_l \delta_{(i-1)-l} ~ a_{l-j} = a_{i-1-j} 	,	\\
	\lb   C_a S \rb_{ij}	& = \sum_l \lb C_a \rb_{il}  S_{lj}  = \sum_l  a_{i-l}  ~\delta_{l-j-1} =  \sum_l  a_{i-l}  ~\delta_{l-(j+1)} = a_{i-j-1} ,
\end{align*}
where~\req{circ_def_ind} is used for the entries of $C_a$. Thus $S$ commutes with any circulant matrix. The converse is also true (see Exercise~\cref{circ_shift.ex}), and we state these conclusions in the next lemma.  
\begin{lemma} 
	A matrix $M$ is circulant iff it commutes with the circular shift operator $S$, i.e. $SM=MS$. 
\end{lemma} 
\begin{figure}
		\centering
		\includegraphics[width=0.7\textwidth]{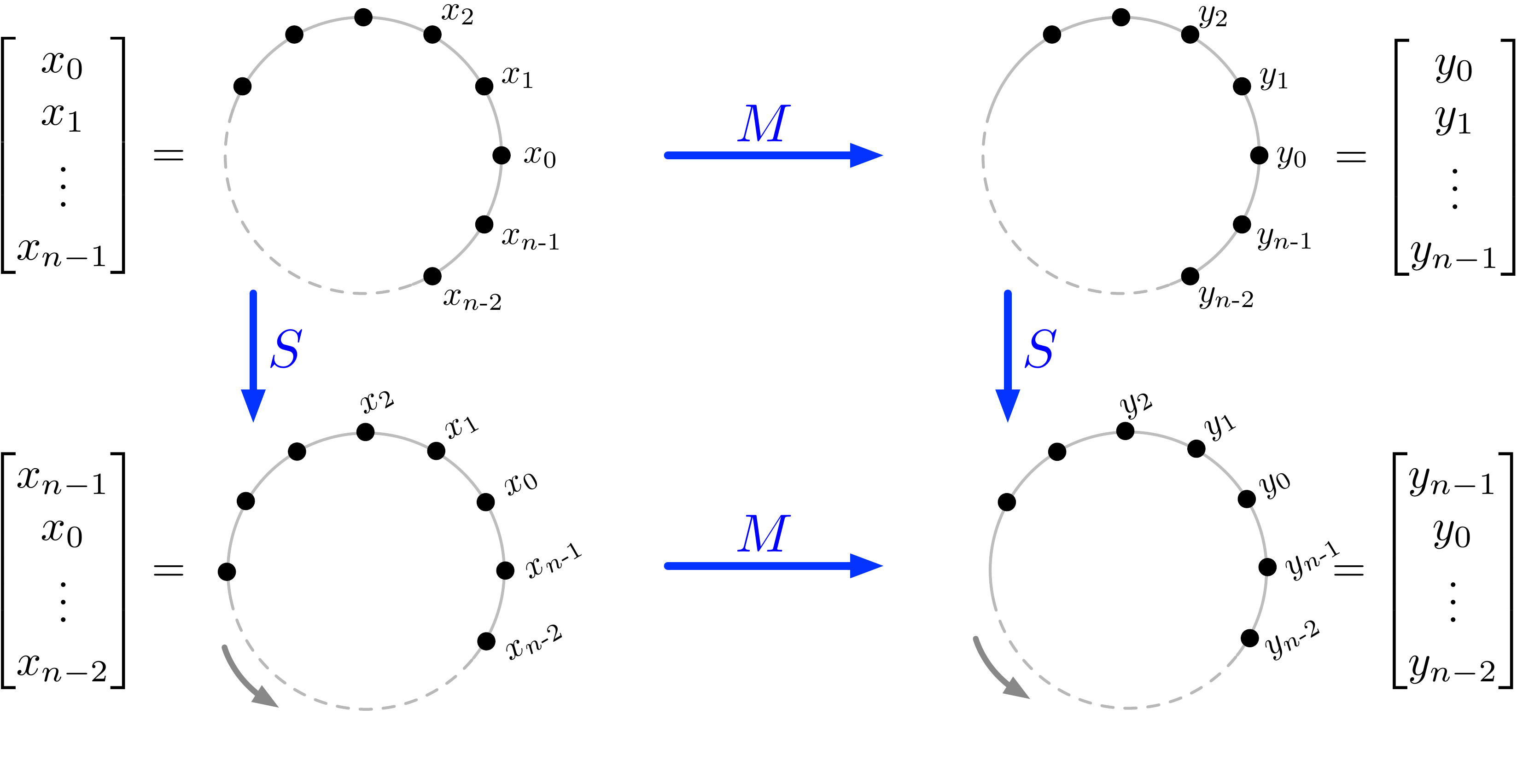}	
	\mycaption{Illustration of the {\em circular shift-invariance} property of the matrix-vector product $y=Mx$ in a commutative diagram. $Sx$ is the circular shift of a vector, depicted also as a counter-clockwise rotation of the vector components arranged on the discrete circle. A matrix $M$ has the shift invariance property if $SM=MS$. In this diagram, this means that the action of $M$ on the rotated vector $Sx$ is equal to acting  on $x$ with $M$ (to produce $y=Mx$) first, and then rotating the resulting vector to yield $Sy$. A matrix $M$ has this shift-invariance property iff it is circulant.  } 
  \label{shift_inv.fig}	 
\end{figure}
Note a simple corollary that a matrix is circulant iff it commutes with $S^*$ since 
\[
	SM ~=~MS 
	~~~~\Longleftrightarrow~~~~ 
	S^* ~SM~ S^* ~=~ S^* ~MS ~ S^*
	~~~~\Longleftrightarrow~~~~ 
	M S^* ~=~ S^* M, 
\]
which could be an alternative statement of the Lemma. 
The fact that a circulant matrix commutes with $S$ could have been used as a definition of a circulant matrix, with the structure in~\req{circ_def} derived as a consequence. Commutation with $S$ also expresses a {\em shift invariance} property. If we think of an $n$-vector $x$ as a function on $\Z_n$ (\cref{circ_func.fig}.b), then $SM x= MSx$ means that the action of $M$ on $x$ is shift invariant. Geometrically, $Sx$ is a counter-clockwise rotation of the function $x$ in~\cref{circ_func.fig}.b. $S\lb Mx\rb = M\lb Sx\rb$ means that rotating the result of the action of $M$ on $x$ is the same as rotating $x$ first and then acting with $M$. This property is illustrated graphically in~\cref{shift_inv.fig}.

\subsection{Circular Convolution}

We will start with examining the matrix-vector product when the matrix is circulant. By analyzing this product, we will obtain the circular convolution of two vectors. Let $C_a$ by some circulant matrix, and examine the action of  such a matrix on any vector $x=\lb x_0 ,~ x_1 ,~ \cdots ,~ x_{n-1} \rb$. The matrix-vector multiplication $y=C_a x$  in detail reads 
\be
	y ~= 	\bbm		 y_0 \\ y_1 \\ \vdots \\ y_{n-1} 	\ebm
	 ~ = ~ 
		 \bbm
			a_0 & a_{n-1} & \cdots & a_{1}	\\
			a_{1} & a_{0} &               & a_{2}	\\
			\vdots     &  &     \ddots           & \vdots	\\
			a_{n-1}    & a_{n-2} &    \cdots           & a_0		
		\ebm
		\bbm	 x_0 \\ x_1 \\ \vdots \\ x_{n-1} 	\ebm
	 	=
		C_a ~x
  \label{circ_mat_vect_x.eq}		 		.		
\ee
Using  $\lb C_a\rb_{kl}=a_{k-l}$, this  matrix-vector multiplication can be rewritten as
\be
	y_{k} ~=~ \sum_{l=0}^{n-1} \lb C_a \rb_{kl} x_l ~=~ \sum_{l=0}^{n-1}  a_{k-l} ~x_l. 
  \label{circ_conv_calc.eq}
\ee
This can be viewed as an operation on the two vectors $a$ and $x$ to yield the vector $y$, and allows us to reinterpret the
matrix-vector product of a circulant matrix as follows. 
\begin{definition} 
Given two $n$-vectors $a$ and $x$, their {\em circular convolution} $y = a\star x$ is another $n$-vector  defined by
\be
	y ~ = ~ a ~\star~ x 	~~~~~~\Leftrightarrow~~~~~~~
	y_k ~ = ~ \sum_{l=0}^{n-1} a_{k-l} ~x_{l},
   \label{circ_conv.eq}
\ee
where the indices in the sum are evaluated modulo $n$. 
\end{definition} 
Comparing~\req{circ_conv_calc} with~\req{circ_conv},
we see that multiplying a vector by a circulant matrix is  equivalent to convolving the vector with the vector defining the 
circulant matrix
\be
	y ~=~ C_a ~x ~=~ a\star x. 
\ee

\begin{figure}
	\begin{center}	\includegraphics[width=0.9\textwidth]{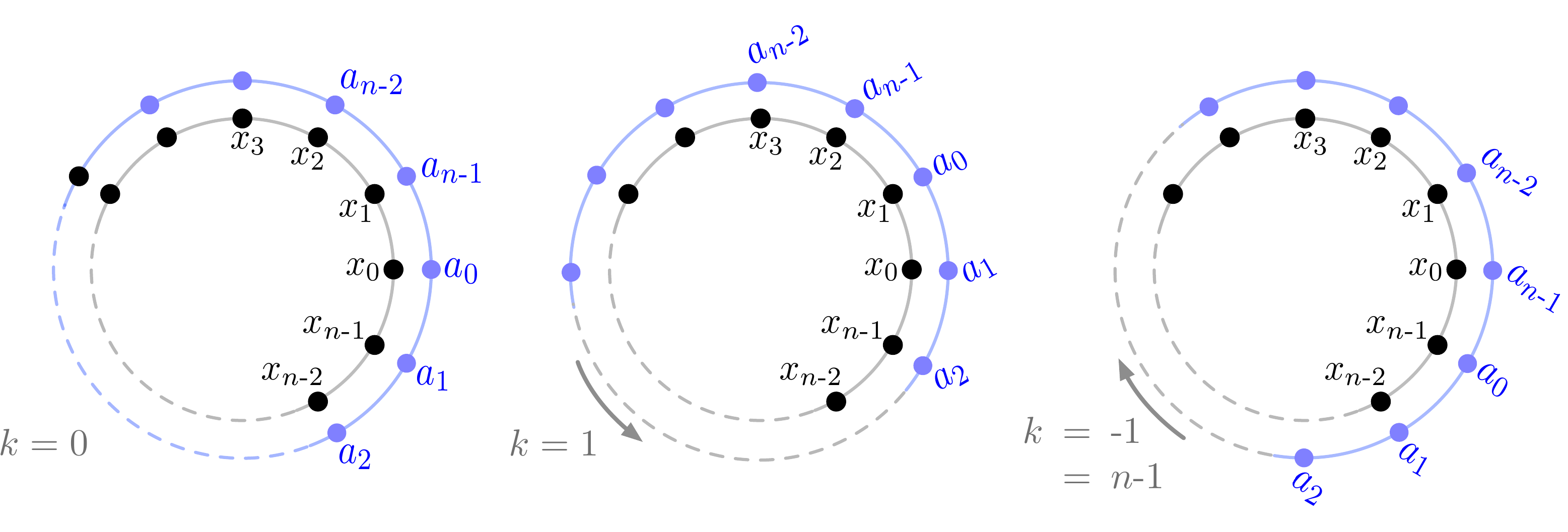}	\end{center}
	\mycaption{Graphical illustration of circular convolution $y_k = \sum_{l=0}^{n-1} \tcb{a_{k-l}} x_l $ for $k=0,1,-1$ respectively. The \tcb{$a$-vector} is arranged in reverse orientation, and then each $y_k$ is calculated from the dot product of $x$ and the \tcb{ rotated, reverse-oriented $a$-vector} rotated by $k$ steps counter clockwise. }
	\label{circ_conv.fig}
\end{figure}

The sum in~\req{circ_conv} defining circular convolution has a nice circular visualization due to modular arithmetic on $\Z_n$. This is   
illustrated in~\cref{circ_conv.fig}. 
The elements of $x$ are arranged in a discrete circle counter-clockwise,
while the elements of $a$ are arranged in a circle clockwise (the reverse orientation is 
because elements of $x$ are indexed like $x_l$ while those of $a$ are indexed like
$a_{.-l}$ in the definition~\req{circ_conv}). For each $k$, the array $a$ is rotated counter-clockwise 
by $k$ steps (\cref{circ_conv.fig} shows  cases for three different values of  $k$). 
The number $y_k$ in~\req{circ_conv} is then
obtained by multiplying the $x$ and rotated $a$ arrays element-wise, 
and then summing. This generates the $n$ numbers $y_0, \ldots, y_{n-1}$. 

From the definition, it is easy to show (see Exercise~\cref{conv_ass_com.ex}) that circular convolution is associative and commutative. 
\begin{itemize} 
	\item \textit{Associativity:} for any three $n$-vectors $a$, $b$ and $c$ we have 
		\[ 	a  \star \lb b \star c \rb ~=~ \lb a \star b \rb \star c 		\] 
	\item \textit{Commutativity:} for any two $n$-vectors $a$ and $b$ 
		\[ 	a \star b ~=~ b \star a		\] 
\end{itemize} 	
The above  two facts have several interesting implications. First, since convolution is commutative, the matrix-vector product~\req{circ_mat_vect_x} can be written in two equivalent ways 
\[
	C_a ~x ~=~  a\star x ~=~x\star a ~=~ C_x ~a. 
\]
Applying this fact  in succession to two circulant matrices
\[
	C_b C_a ~x ~=~ C_b \lb a\star x \rb ~=~  b \star \lb a \star x \rb  
		~=~  \lb b \star  a \rb \star x   
		~=~ C_{b\star a} ~x .
\]
This means that the product of any two circulant matrices $C_b$ and $C_a$ is another circulant matrix $C_{b\star a}$ whose defining vector is $b\star a$, the circular convolution of the defining vectors of $C_b$ and $C_a$ respectively. We summarize this conclusion and an important corollary of it next. 
\begin{theorem}
\begin{enumerate}
		\item Circular convolution of any two vectors can be written as a matrix-vector product with a circulant matrix 
			\[
				a\star x ~=~ C_a ~x ~=~ C_x ~a. 
			\]
		\item The product  of any two circulant matrices is another circulant matrix 
			\[	C_a C_b ~=~ C_{a\star b}.		\]
		\item All circulant matrices mutually commute since for any two $C_a$ and $C_b$ 
			\[
				C_a C_b ~=~ C_{a\star b} ~=~ C_{b\star a} ~=~ C_b C_a . 
			\]
\end{enumerate}
\end{theorem}

The set of all $n$-vectors forms a commutative algebra under the operation of circular convolution. The above shows that the 
set of $n\times n$ circulant matrices under standard matrix multiplication is also a commutative algebra isomorphic to $n$-vectors with circular convolution.

\section{Simultaneous Diagonalization  of all Circulant Matrices Yields the DFT }					\label{sdiag.sec}

In this section, we will {\em derive} the DFT as a byproduct of diagonalizing circulant matrices. Since all circulant matrices mutually commute, we recall~\cref{s_diag.lemma} and look for a circulant matrix that has simple eigenvalues. The eigenvectors of that matrix will then give the simultaneously diagonalizing transformation. 

The shift operator is in some sense the most fundamental circulant matrix, and is therefore a good candidate for an eigenvector/eigenvalue decomposition. The eigenvalue problem for $S$ will turn out to be the simplest one.  
Note that we have two options. To find eigenvectors of $S$ or alternatively of $S^*$. We begin with $S^*$ since this will end up yielding the classically defined DFT.

\subsection{Construction of Eigenvectors/Eigenvalues of $S^*$} 

Let  $w$ be an eigenvector (with eigenvalue $\lambda$) of the shift operator $S^*$.  Note  that it is also an eigenvector (with eigenvalue $\lambda^l$) of any power $(S^*)^l$ of $S^*$. Applying the definition~\req{S_ind} to the  relation $S^*w=\lambda w$ will reveal that  an eigenvector $w$ has a very  special structure
\be
	\begin{array}{rclcrcll}
            	S^* w &=&  \lambda w
            	&~~~~~\Longleftrightarrow~~~~~&
            	 w_{k+1} 	&=&   \lambda~ w_k, 
            				&	~~~~~k\in\Z_n	,	\\
           	(S^*)^l w &=&  \lambda^l w
            	&~~~~~\Longleftrightarrow~~~~~&
            	 w_{k+l} 	&=&   \lambda^l~ w_k, 
            				&	~~~~~k\in\Z_n	, ~l\in\Z,
	\end{array}
   \label{egvl_circ.eq}
\ee
i.e. each entry $w_{k+1}$ of $w$ is equal to the previous entry $w_k$ multiplied by the eigenvalue $\lambda$. 
These relations can be used to compute all eigenvectors/eigenvalues  of $S^*$. First, observe that although~\req{egvl_circ} is valid for all $l\in\Z$, this relation ``repeats'' for $l\geq n$. In particular, for $l=n$ we have for  each index $k$
\be
	w_{k+n} ~=~ \lambda^n w_k 
	~~~~\Longleftrightarrow~~~~
	w_{k} ~=~ \lambda^n w_k 
  \label{lambda_discovery.eq}
\ee
since $k+n \equiv_n k$. Now since the vector $w\neq 0$, then for at least one index $k$, $w_k\neq 0$, and the last equality implies that  
$
	\lambda^n = 1, 
$
i.e. any eigenvalue of $S$ must be an {\em $n$th root of unity }
\[
	\boxed{
	\lambda^n = 1 
	~~~~~\Longleftrightarrow~~~~~
	\lambda ~=~ \rho_m ~:=~  e^{i\frac{2\pi}{n} m } , ~~m\in\Z_n.		}
\]
{\em Thus we have discovered that the $n$ eigenvalues of $S^*$ are precisely the $n$ distinct $n$th roots of unity $\lcb \rho_m, ~m=0,\ldots,n-1 \rcb$. }
Note that any of the $n$th roots of unity can be expressed as a power of the first $n$th root: $\rho_m = \rho_1^m$ (recall~\cref{Z_n:roots.fig}). 

\begin{figure}[h]
		\centering		\newcommand{\sfh}{0.15}			
		\setlength{\tabcolsep}{1pt}				
		\begin{tabular}{cccc}
			\begin{tabular}{c}
				\includegraphics[height=\sfh\textheight]{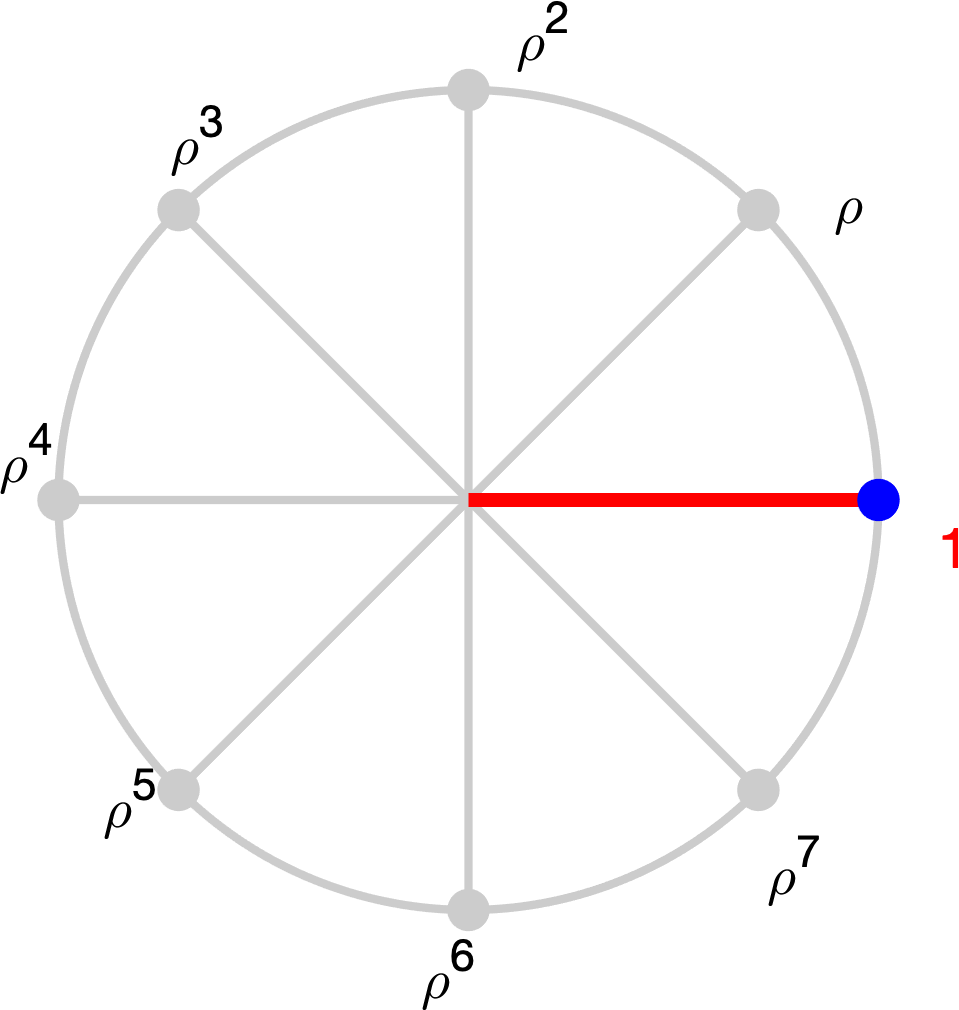}	 \\
				\small \colb $w^{(0)} = $		\vspace*{-.6em} \\ 	
				\tiny $ (0,0,0,0,0,0,0,0)$ 
			\end{tabular} 
		&
			\begin{tabular}{c}
				\includegraphics[height=\sfh\textheight]{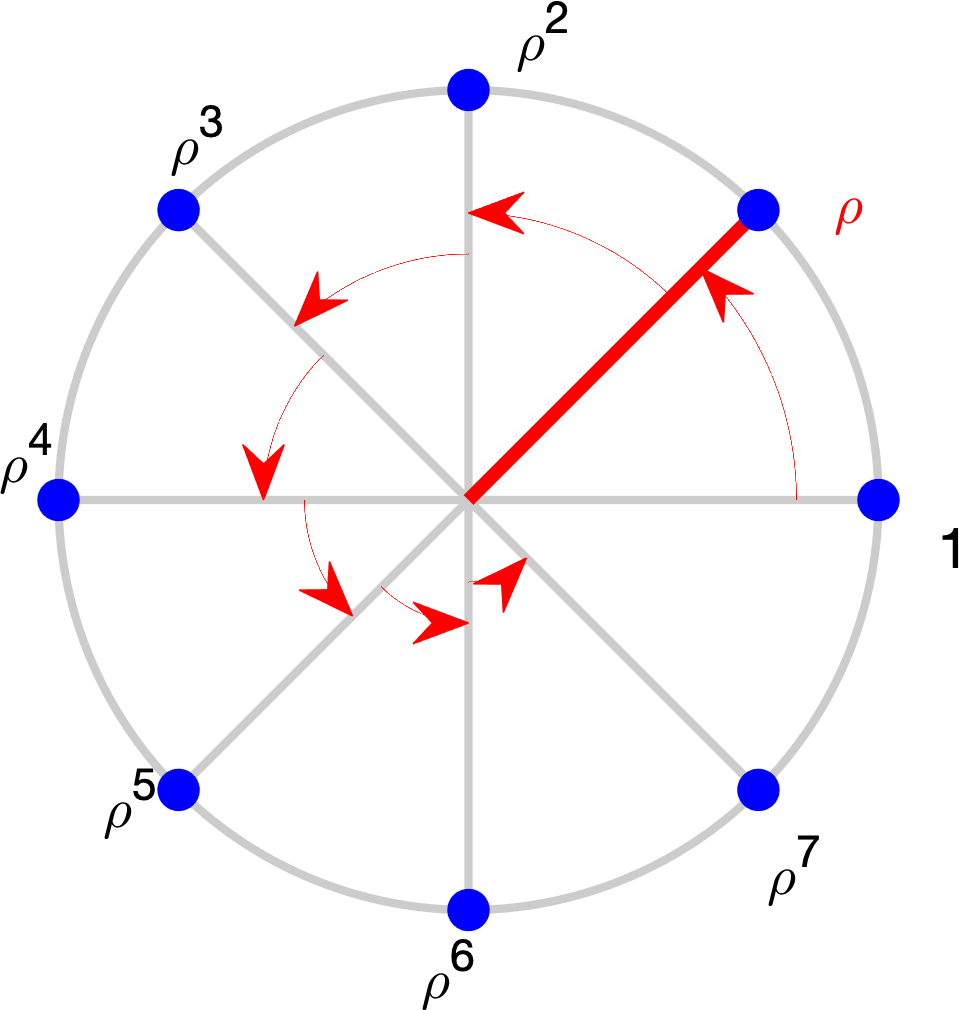}	\\
				$w^{(1)} = $ 		\vspace*{-.6em}\\ 
				\tiny $ (0,1,2,3,4,5,6,7)$ 	
			\end{tabular} 			
		&
			\begin{tabular}{c}
				\includegraphics[height=\sfh\textheight]{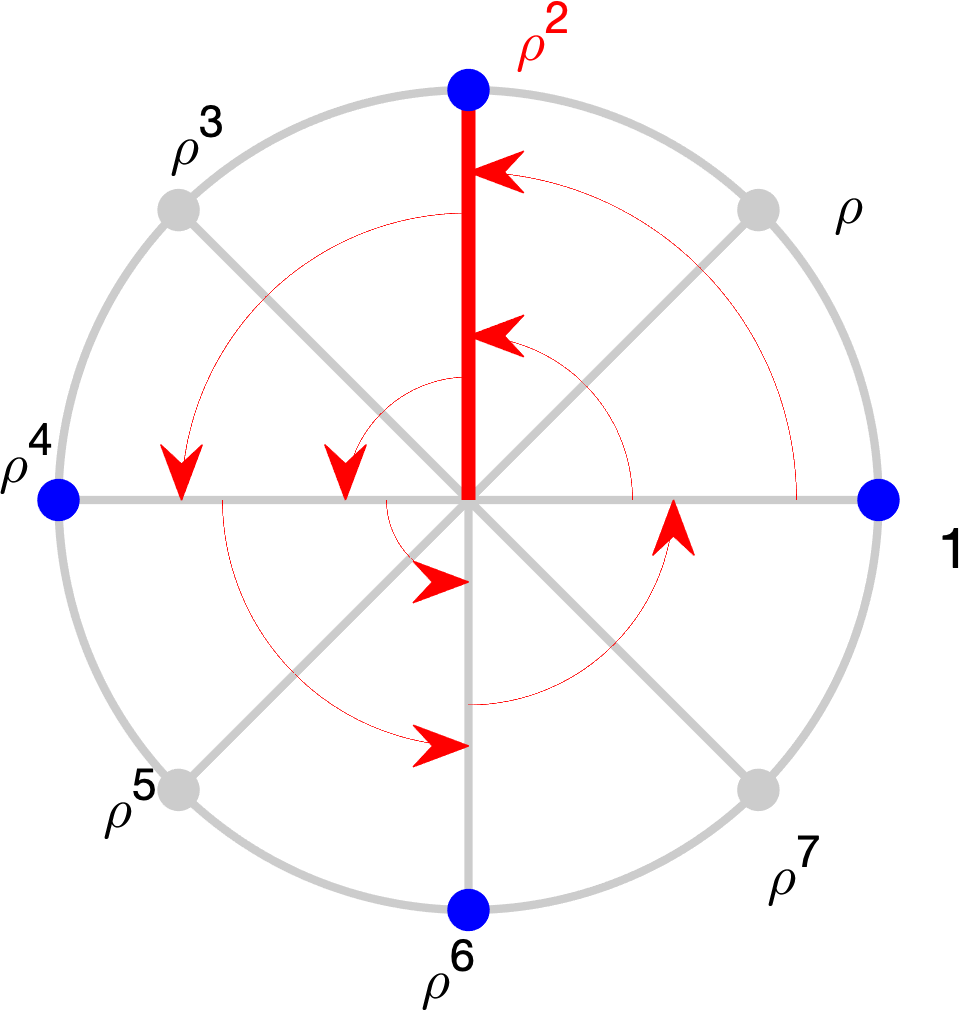}	\\
				$w^{(2)} = $ 		\vspace*{-.6em}\\ 
				\tiny $ (0,2,4,6,0,2,4,6)$ 	
			\end{tabular} 			
		&
			\begin{tabular}{c}
				\includegraphics[height=\sfh\textheight]{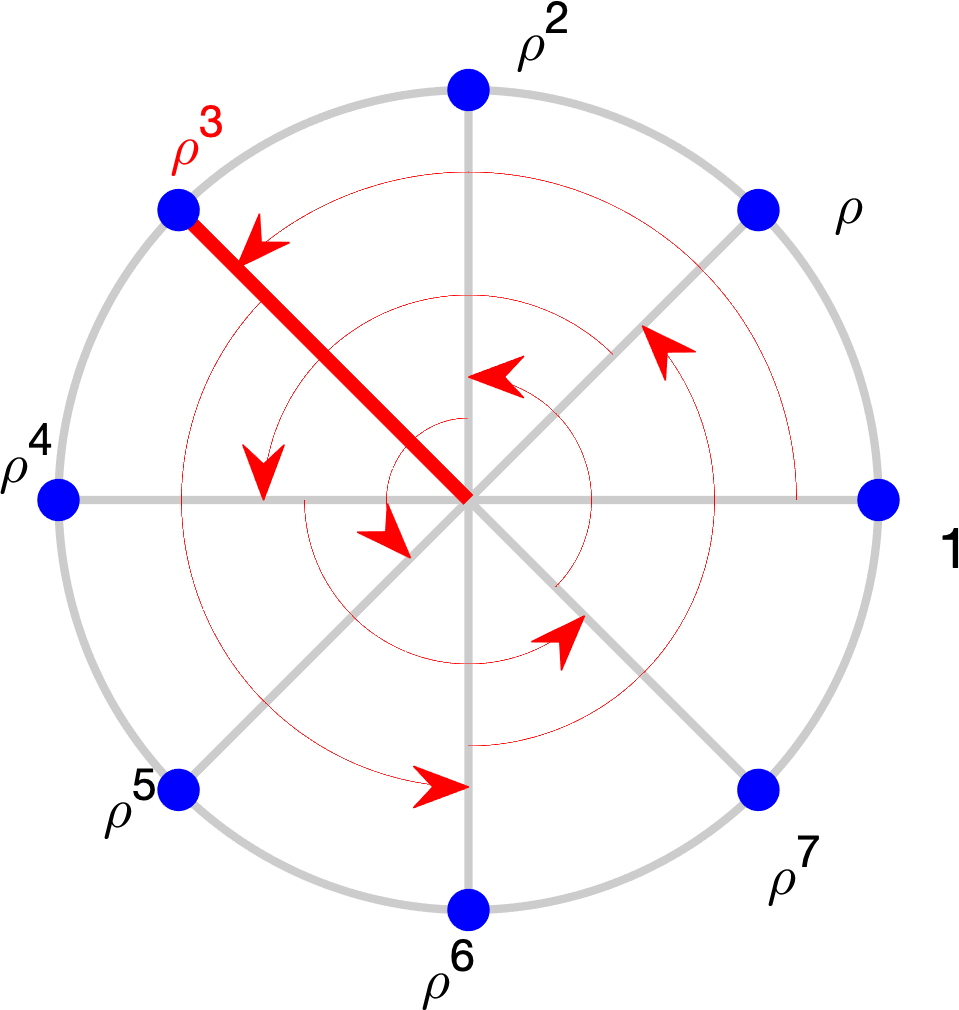}	\\
				$w^{(3)} = $ 		\vspace*{-.6em}\\ 
				\tiny $ (0,3,6,1,4,7,2,5)$ 	
			\end{tabular} 			
		\\		\\
			\begin{tabular}{c}
				\includegraphics[height=\sfh\textheight]{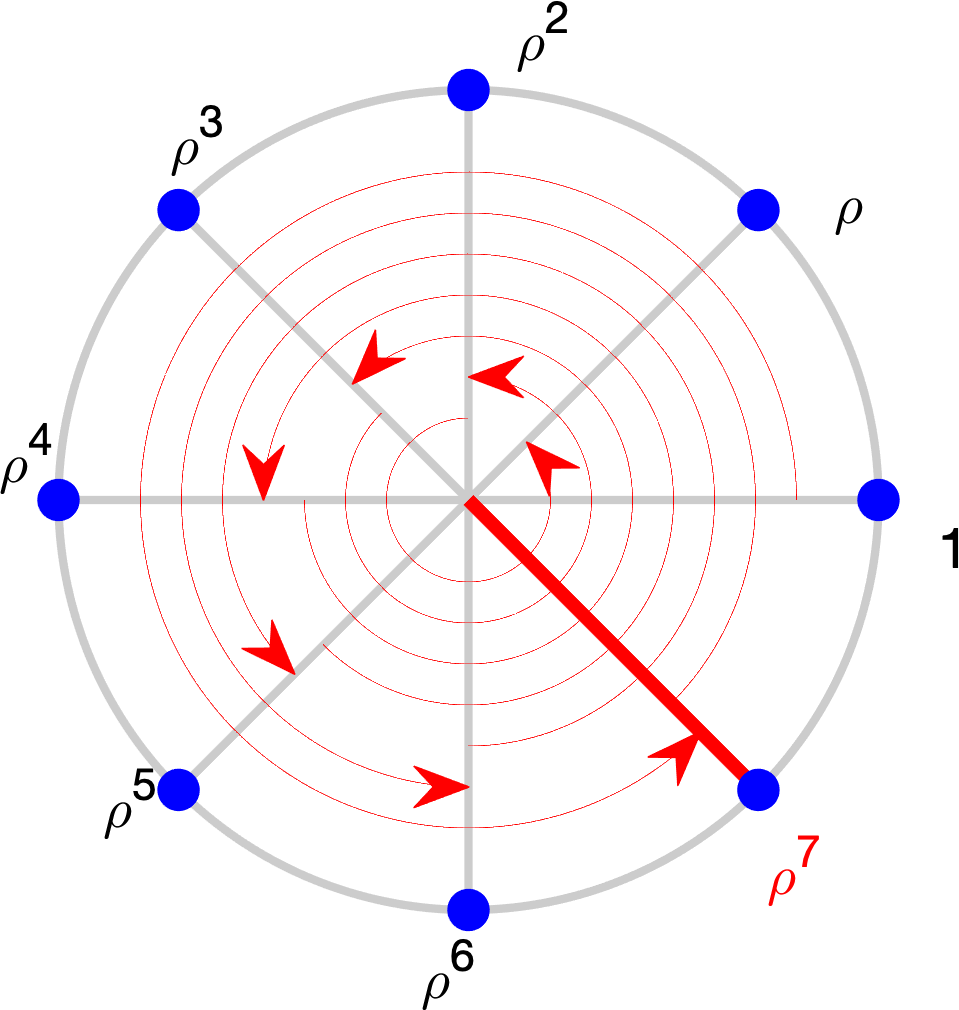}	\\
				$w^{(7)} = $		\vspace*{-.6em} \\ 
				\tiny $ (0,7,6,5,4,3,2,1)$ 	
			\end{tabular} 			
		&
			\begin{tabular}{c}
				\includegraphics[height=\sfh\textheight]{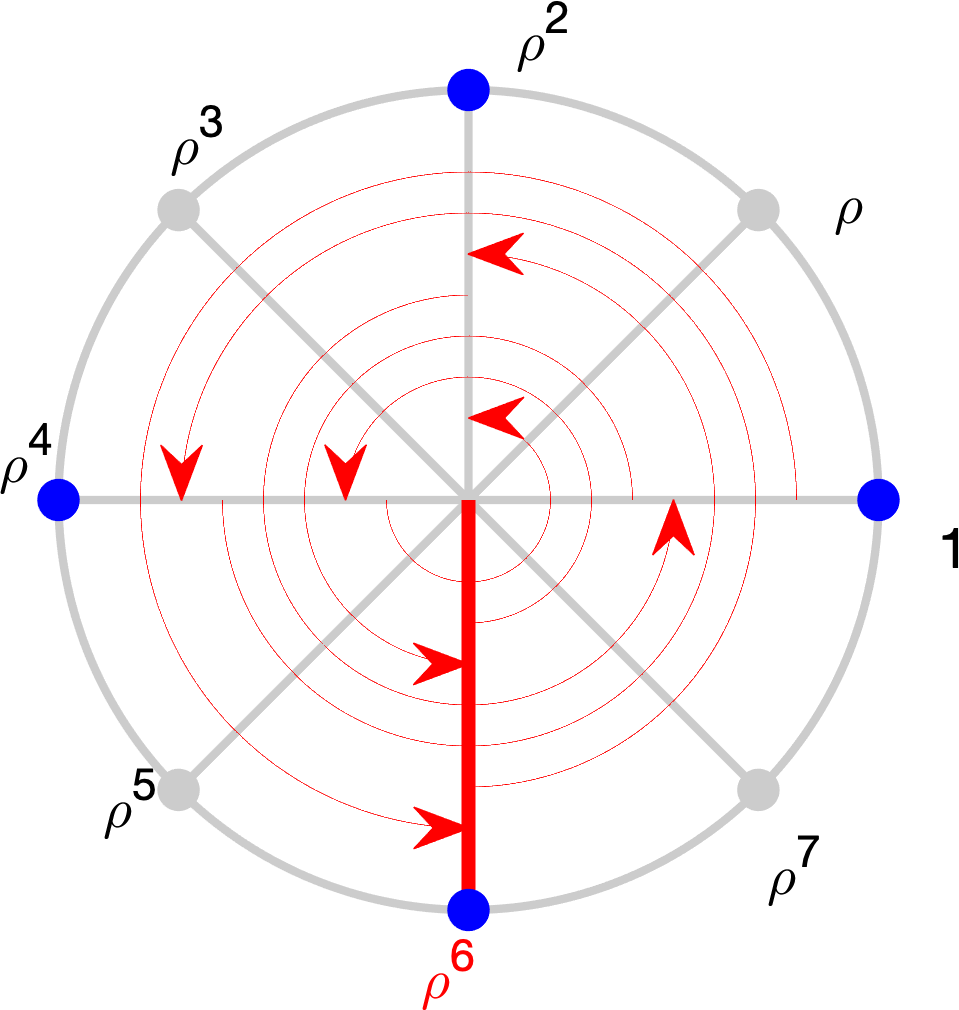}	\\
				$w^{(6)} = $		\vspace*{-.6em} \\ 
				\tiny $ (0,6,4,2,0,6,4,2)$ 	
			\end{tabular} 			
		&
			\begin{tabular}{c}
				\includegraphics[height=\sfh\textheight]{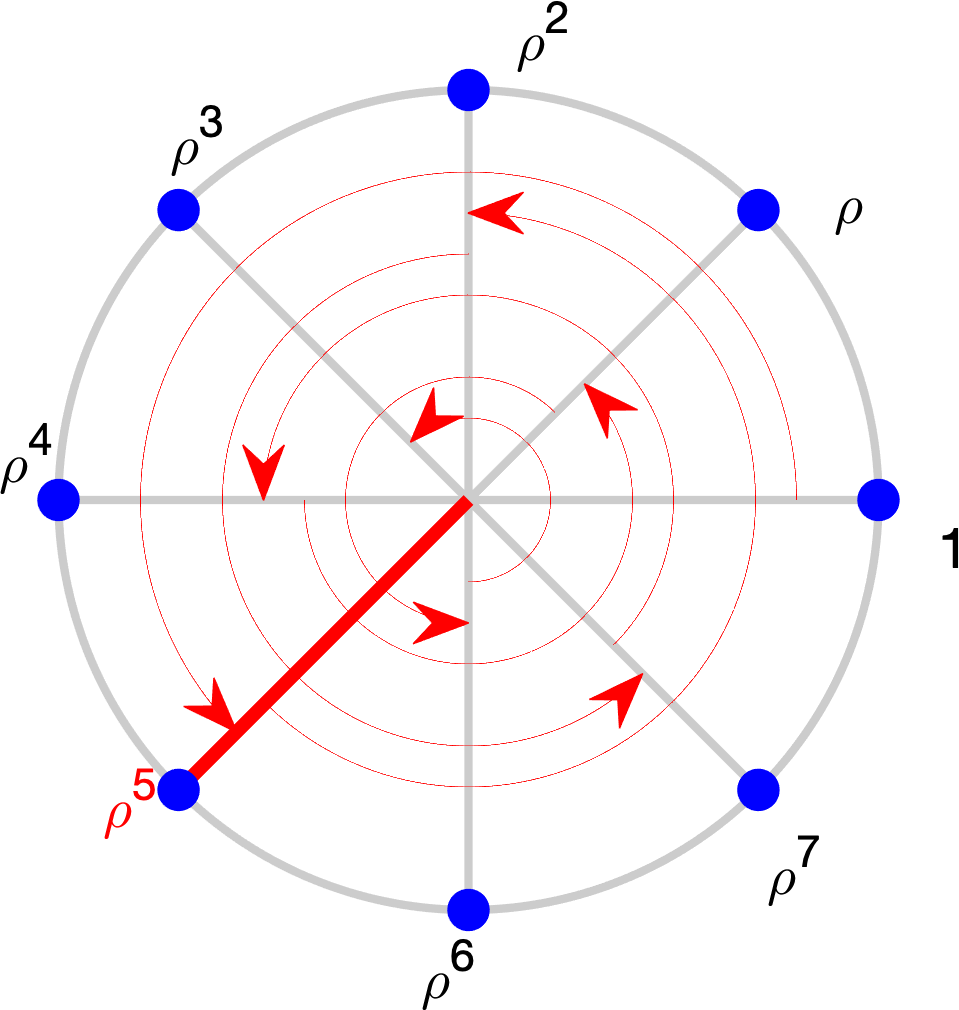}	\\
				$w^{(5)} = $		\vspace*{-.6em} \\ 
				\tiny $ (0,5,2,7,4,1,6,3)$ 	
			\end{tabular} 			
		&
			\begin{tabular}{c}
				\includegraphics[height=\sfh\textheight]{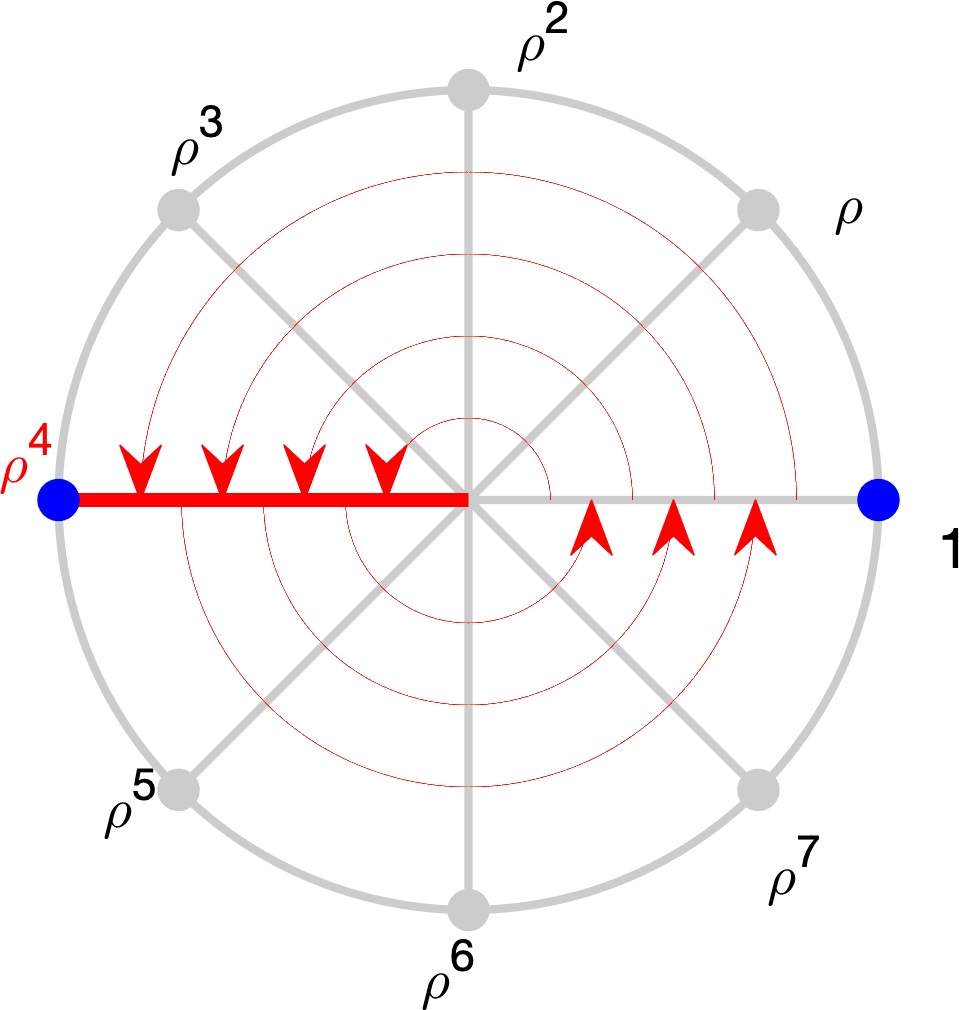}	\\
				$w^{(4)} = $ 		\vspace*{-.6em}\\ 
				\tiny $ (0,4,0,4,0,4,0,4)$ 	
			\end{tabular} 			
		\end{tabular} 			
	\mycaption{Visualization of the eigenvalues and eigenvectors of the left shift operator $S^*$ for the case $n=8$. 
			The eigenvalues (red straight lines and red labels), 
			and elements of the corresponding eigenvectors (blue dots) 
			are all points on the unit circle of the complex plane. 
			$\rho$ is the $n$'th root of unity (here $\rho=e^{i{\pi/4}}$).  For each $m\in\Z_n$, 
			$\rho^m$ is an eigenvalue with eigenvector
			 $w^{(m)}=\lb 1, \rho^m, \rho^{2m}, \ldots,\rho^{m(n-1)} \rb$, where each element is a 
			 rotation of the previous element by $\rho^m$ (curvy red arrows). For compactness of notation, {\em vectors
			  are denoted above by powers of $\rho$, e.g. $w^{(4)}=\lb 1, \rho^2, \rho^4, \rho^6, 1, \rho^2, \rho^4, \rho^6 \rb
			  = (0,2,4,6,0,2,4,6)$}. Notice the pattern that which powers of $\rho$ appear in $w^{(m)}$ depends on the 
			  least common factor (lcf) of $m$ and $n$, e.g. in $w^{(4)}$ that number is $8/{\rm lcf}(4,8)~=~ 2$. 
			  For $(m,n)$ co-prime, all powers of $\rho$ appear in $w^{(m)}$, though with permuted ordering (see 
			  $w^{(1)}, w^{(3)}, w^{(5)}, w^{(7)}$). 
			 } 
  \label{eigs_viz.fig}	 
\end{figure}

Now fix $m\in\Z_n$ and compute $w^{(m)}$,  the eigenvector corresponding to the eigenvalue $\rho_m$. Apply the last relation in~\req{egvl_circ}  $w_{k+l}=\lambda^{ l} w_k$, and use it to express the entries of the eigenvector $w^{(m)}$  in terms of the first entry ($k=0$)
\be
	w^{(m)}_{l+0} = \lambda^{l} w_0 
	~~~~\Leftrightarrow~~~~
	w^{(m)}_{l} =  \rho_m^l  w_0
	~~~~\Leftrightarrow~~~~
	w^{(m)}  =   w_0 \lb 1 , ~\rho_m , ~\rho_m^{ 2} ,~ \ldots , ~\rho_m^{n\sm 1} \rb .
   \label{w_discovery.eq}	
\ee
Note that $w_0$ is a scalar, and since eigenvectors are only unique up to multiplication by a scalar, we can  set $w_0=1$ for a more compact expression for the eigenvector. In addition, $\rho_m$ in~\req{w_discovery} could be any of the $n$th roots of unity, and thus that expression applies to all of them, yielding the $n$ eigenvectors.  
We summarize the previous derivations in the following statement. 
\begin{lemma} 											\label{S_eig.lemma}
	The circular left-shift operator $S^*$ on $\R^n$ has  $n$ distinct eigenvalues. They are the $n$th roots of unity $\rho_m := e^{i\frac{2\pi}{n} m} = \rho_1^m =:\rho^m $, $m\in\Z_n$. The corresponding  eigenvectors are 
	\be
		w^{(m)} ~=~  \lb 1 ,~ \rho^{m} ,~ \rho^{ 2m} ,~  \ldots ,~ \rho^{m(n\sm 1)} \rb , 
		~~~~~~~~~~m = 0,\ldots,n-1, 
	   \label{S_eigenvec.eq}
	\ee
\end{lemma} 
Note that the  eigenvectors $\lcb w^{(m)} \rcb$ are indexed with the same index as the eigenvalues $\lcb \lambda_m = \rho_m \rcb$.
It is useful and instructive to visualize the eigenvalues and their corresponding eigenvectors as specially ordered sets of the 
roots of unity. Which roots of unity enter into any particular eigenvector, as well as their ordering, is determined by the algebra of 
rotations of roots of unity.    This is illustrated in  detail in~\cref{eigs_viz.fig}


\subsection{Eigenvalues Calculation  of a Circulant Matrix Yields the DFT} 

Now that we have calculated all the eigenvectors of the shift operator in~\cref{S_eig.lemma}, we can use them to find the eigenvalues of any circulant matrix $C_a$. Recall that since any circulant matrix commutes with $S^*$, and $S^*$ has distinct eigenvalues,   then $C_a$ 
has the same eigenvectors as those~\req{S_eigenvec} previously found for $S^*$ (by~\cref{s_diag.lemma}). Thus we have  the relation
\be
	C_a ~w^{(m)} = \lambda_m ~w^{(m)} 
	~~~~\Leftrightarrow~~~~
 	\bbm	a_0 		& a_{n-1}	& \cdots  & a_{1}	\\
			a_{1} 	& a_{0} 	&              & a_{2}	\\
			\vdots	&  		& \ddots 	& \vdots	\\
			a_{n-1}    & a_{n-2} &  \cdots	& a_0	\ebm
	\bbm	1   \\   \rho_m \\ \vdots 	\\ \rho_m^{n\sm 1} 	\ebm
	= \lambda_m
	\bbm	1   \\   \rho_m \\ \vdots 	\\ \rho_m^{n\sm 1} 	\ebm ,
  \label{circ_eigen.eq}
\ee
where $\{\lambda_m\}$ are the eigenvalues of $C_a$ (not the eigenvalues of $S^*$ found in the previous section). 
 Each row of the above equation represent essentially the same equation (but multiplied by a power of $\rho_m$). The first row is the easiest equation to work with 
\begin{align}
	\hspace*{-1em}
	\lambda_m 
			   &=~ a_0 + a_{n-1}~ \rho_m +  \cdots  + a_1~ \rho_m^{n\sm 1}					\nonumber	\\ 
			   &=~ a_0  + a_1~ \rho_m^{\sm 1}  +  \cdots  	+ a_{n-1}~ \rho_m^{\sm (n\sm 1)}	\nonumber	\\
			   &=~ \sum_{l=0}^{n-1} a_l ~\rho_m^{- l} 
			   ~=~ \sum_{l=0}^{n-1} a_l ~\rho^{-m l} 
			   ~=~ \boxed{  \sum_{l=0}^{n-1} a_l ~e^{- i\frac{2\pi}{n} ml }  ~=:~ \ah_m, }
  \label{DFT_discovery.eq}			   
\end{align}
which is precisely the classically-defined DFT~\req{DFT_def} of the vector $a$. 

We therefore conclude that any circulant matrix $C_a$ is diagonalizable by the basis~\req{S_eigenvec}. Its $n$ eigenvalues are given by $\lb \ah_0, \ah_1, \ldots, \ah_{n-1} \rb$ from~\req{DFT_discovery}, which is  the DFT of the vector $\lb a_0, a_1, \ldots, a_{n-1}\rb$. In this way, the DFT arises from a formula for computing the eigenvalues of any circulant matrix. 


One might ask what the conclusion would have been if the eigenvectors of $S$ have been used instead of those of $S^*$. A repetition of the previous steps but now for the case of $S$ would yield that the eigenvalues of a circulant matrix $C_a$ are given by 
\be
	\mu_k ~=~ \sum_{l=0}^{n-1} a_l ~e^{i\frac{2\pi}{n} kl} , ~~~~~~~~~ k=0,1,\ldots, n-1.
   \label{DFT_alternate.eq}	
\ee
While the expressions~\req{DFT_discovery} and~\req{DFT_alternate} may at first appear different, the sets of numbers $\lcb \lambda_m \rcb$ and $\lcb \mu_k \rcb$ are actually equal. So in fact, the expression~\req{DFT_alternate} gives the same set of eigenvalues as~\req{DFT_discovery} but arranged in a different order since $\mu_k = \lambda_{-k}$. 
\[
	\lambda_{-k} ~=~ \sum_{l=0}^{n-1} a_l ~e^{- i\frac{2\pi}{n} (-k)l }~=~ \sum_{l=0}^{n-1} a_l ~e^{i\frac{2\pi}{n} kl } ~=~ \mu_k.
\]

Along with the two choices of $S$ and $S^*$, there are also other possibilities. Let $p$ be any number that is coprime with $n$. It is easy to show (Exercise~\cref{p_prime.ex}) that a $n\times n$ matrix is circulant iff it commutes with $S^p$. In addition, the eigenvalues of $S^p$ are distinct (see~\cref{eigs_viz.fig}).  Therefore the eigenvectors of $S^p$ (rather than those of $S$) can be used to simultaneously diagonalize all circulant matrices. This would yield yet another transform distinct from the two transforms~\req{DFT_discovery} or~\req{DFT_alternate}. However, the set of numbers produced from that transform will still be the same as those computed from the previous two transforms, but arranged in a different ordering.

\section{The Big Picture}														\label{big.sec}

Let $C_a$ be a circulant matrix made from a vector $a$ as in~\req{circ_def}. If we use the eigenvectors~\req{S_eigenvec}  of $S^*$ as columns of a matrix $W$, the $n$ eigenvalue/eigenvector relationships~\req{circ_eigen} $C_a w^{(m)} = \lambda_m w^{(m)}$  can be written as a single matrix equation as follows 
\be
	\begin{aligned}
	C_a
		\matbegin \!\!\! \begin{array}{c:c:c}		
       				w^{(0)}   \rule[-1.7em]{0em}{4em}  & \cdots &  w^{(n\sm 1)}
       				\end{array} \!\!\!    \matend 
		& = 
		\matbegin	\!\!\!	 \begin{array}{c:c:c}
       				w^{(0)}   \rule[-1.7em]{0em}{4em}  & \cdots &  w^{(n\sm 1)}
       				\end{array} 	\!\!\!	 \matend 
			\matbegin	\!\!	 \begin{array}{ccc}
       				\ah_0 & & \\
				 & \!\!\! \smash{ \ddots} \!\!\! & \\
				  & &\ah_{n-1}
       				\end{array}	\!\! \matend  ,	\\
	\Longleftrightarrow~~~~~~~~~~~~
	C_a~ W& =~   W ~\diago{\ah} ,
	\end{aligned}
  \label{Cx_W.eq}	
\ee
where we have used the fact~\req{DFT_discovery} that the eigenvalues  of $C_a$ are precisely $\lcb \ah_m\rcb$, the elements of the DFT of the vector $a$. 

It is easy to verify that the columns of $W$ are mutually orthogonal\footnote{
	This also follows from the fact that the columns of $W$ are the eigenvectors of $S^*$, and since $S^*$  is a normal matrix, it has mutually orthogonal eigenvectors. 
}, 
and thus $W$ is a unitary matrix (up to a rescaling) $W^*W=WW^*=nI$, or equivalently $W^{-1} = \frac{1}{n} W^*$.  
Since the matrix $W$ is made up of the eigenvectors of $S^*$, which in turn are made up of various powers of the roots of unity~\req{S_eigenvec}, it has some special structure which is worth examining
\[
	W:= 
		\matbegin \!\!\! \begin{array}{c:c:c}		
       				w^{(0)}   \rule[-1.7em]{0em}{4em}  & \cdots &  w^{(n\sm 1)}
       				\end{array} \!\!\!    \matend 
		= 	
		\bbm 
				1 		&	1			& 	\cdots	& 	1				\\ 
				1		& 	\rho			&	\cdots	& 	\rho^{n\sm 1}		\\
				\vdots	&	\vdots		&			&	\vdots			\\ 
				1		&	\rho^{n\sm 1} 	&	\cdots	&	\rho^{(n\sm 1) (n\sm 1)}
		\ebm. 
\]
The matrix $W$ is symmetric,  $W^*$ is thus the matrix $W$ with each entry replaced by its complex conjugate. Furthermore, since for each root of unity $\lb\rho^k\rb^*=\rho^{-k}$, we can therefore write 

\[
	W^*= 
		\bbm 
				1 		&	1			& 	\cdots	& 	1				\\ 
				1		& 	\rho^{\text{-1}}			&	\cdots	& 	\rho^{\text{-}(n\sm 1)}		\\
				\vdots	&	\vdots		&			&	\vdots			\\ 
				1		&	\rho^{\sm(n\sm 1)} 	&	\cdots	&	\rho^{\sm (n\sm 1) (n\sm 1)}
		\ebm. 
\]
Also observe that multiplying a vector by $W^*$ is exactly taking its  DFT. Indeed the $m$'th row of $W^*x$ is 
\[
	\xh_m ~=~ 
		\bbm 1 & \rho^{\sm m} & \cdots & \rho^{\sm m(n\text{-1})}		\ebm
		\bbm  x_0  \\ \vdots \\ x_{n-1} \ebm, 
\]
which is exactly the definition~\req{DFT_def} of the DFT. 
Similarly, multiplication by $\frac{1}{n}W$ is taking the inverse DFT 
\[
	x_l ~=~\frac{1}{n}  \sum_{k=0}^{n-1} \xh_k ~\rho^{kl}  
		~=~\frac{1}{n}  \sum_{k=0}^{n-1} \xh_k ~e^{i \frac{2\pi}{n} kl} . 
\]

\begin{figure}[h]
	\centering
	\includegraphics[width=0.6\textwidth]{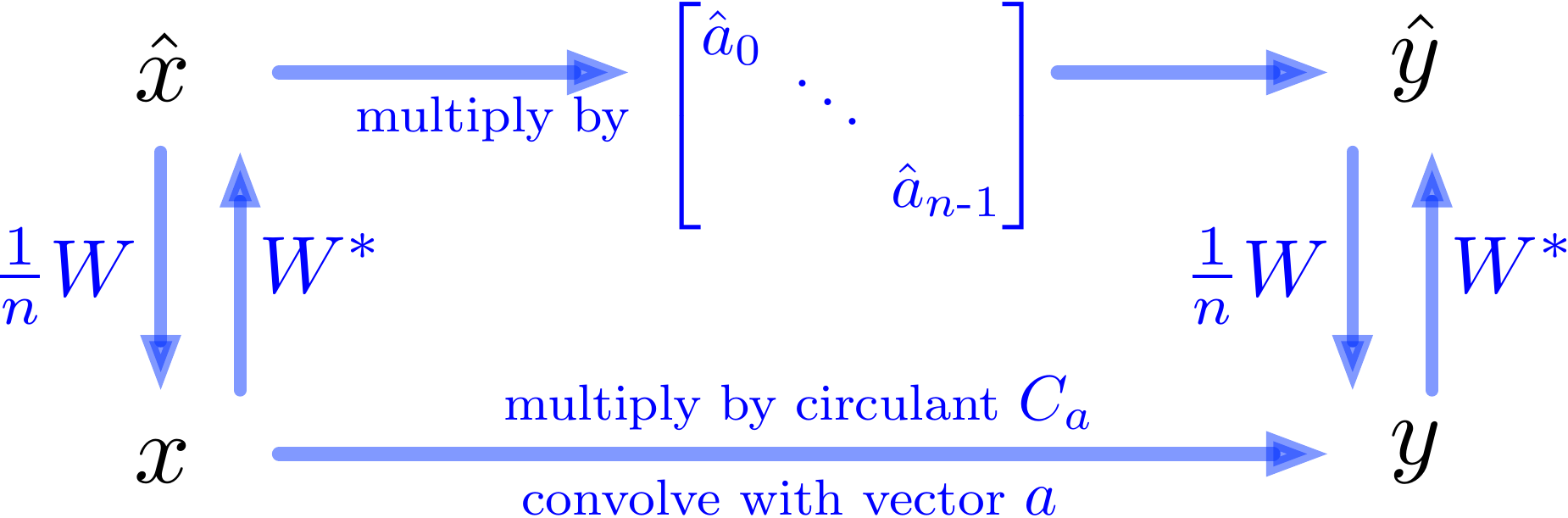}
    \mycaption{Illustration of the relationships between circulant matrices, circular convolution and the DFT. The matrix-vector multiplication $y=C_a x$ with the circulant matrix $C_a$ is equivalent to the circular convolution $y=a\star x$. The DFT is a linear transformation $W^*$ on vectors with inverse $\frac{1}{n}W$. It converts multiplication by the circulant matrix $C_a$ into multiplication by the diagonal matrix $\rm{diag}\lb \ah_0, \ldots, \ah_{n\text{-1}}\rb$ whose entries are the DFT of the vector $a$ defining the matrix $C_a$. }
    \label{commut.fig}
\end{figure}

Multiplying both sides of~\req{Cx_W} from the right by $W^{-1}$ gives 
 the diagonalization of $C_a$ which can be written in several equivalent forms 
\begin{align}
	C_a &=~ W~\diago{\ah}~  W^{-1}
	 ~=~ W~\diago{\ah}~ \lb \mbox{ $\frac{1}{n}$}   W^{*} \rb											\nonumber	\\
	 &=~  \lb \mbox{ $\frac{1}{n}$}   W \rb  ~\diago{\ah}~  W^*										
	 ~=~  \lb \mbox{ $\frac{1}{\sqrt{n}}$}   W \rb  ~\diago{\ah}~   \lb \mbox{ $\frac{1}{\sqrt{n}}$}   W^* \rb .		\label{diagon_1.eq}	
\end{align}
The diagonalization~\req{diagon_1} can be interpreted as follows in
terms of the action of a circulant matrix $C_a$   on any vector $x$ 
\[
	C_a ~x ~=~      \underbrace{    \lb \mbox{ $\frac{1}{n}$}   W \rb   \underbrace{~ \diag{\ah}  ~~~~    \underbrace{~W^*~ x ~}_{ \mbox{\scriptsize DFT of $x$}} ~~}_{\mbox{\scriptsize multiply by $\ah$ entrywise}}}_{\mbox{\scriptsize inverse DFT }}
\] 
Thus the action of $C_a$ on $x$, or equivalently the circular convolution of $a$ with $x$, can be performed by first taking the DFT of $x$, then multiplying the resulting vector component-wise by $\ah$ (the DFT of the vector $a$ defining the matrix $C_a$), and then taking an inverse DFT. In other words,  the diagonalization of a circulant matrix is equivalent to converting circular convolution to component-wise vector multiplication through the DFT. This is illustrated in~\cref{commut.fig}.
%

Note that in the literature there is an alternative form for the DFT and its inverse 
\[
	\xh_k ~=~\frac{1}{\sqrt{n}}  \sum_{l=0}^{n-1} x_l ~e^{-i \frac{2\pi}{n} kl} , ~~~~~~~~~
	x_l ~=~\frac{1}{\sqrt{n}}  \sum_{k=0}^{n-1} \xh_k ~e^{i \frac{2\pi}{n} kl} , 
\]
which is sometimes preferred due to its symmetry (and is also truly unitary since with this definition $\|x\|_2 = \|\xh\|_2$). This ``unitary'' DFT corresponds to the last diagonalization  given in~\req{diagon_1}. We do not adopt this unitary DFT definition here since it 
complicates\footnote{If the unitary DFT is adopted, the equivalent statement would be that the eigenvalues of $C_a$ are the elements of the entries of $\sqrt{n}~ \ah$. 	}
the statement that the eigenvalues of $C_a$ are precisely the entries of $\ah$.  

We summarize the algebraic aspects of the big picture in the  following theorem.

\begin{theorem} 
		 The following sets are isomorphic commutative algebras
		\begin{enumerate}
			\item[(a)]
			The set of $n$-vectors is closed under circular convolutions and is thus an 
			algebra with the operations of addition and convolution.  
			\item[(b)] The set of $n\times n$ circulant matrices is an algebra under the operations
			of addition and matrix multiplication. 
			\item[(c)] The set of $n$-vectors is an algebra under the operations of  
			addition and component-wise multiplication. 
		\end{enumerate} 
		
		The above isomorphisms are depicted by the following diagram
		
		\bigskip
		
		\begin{center} 
			\includegraphics[height=0.13\textheight]{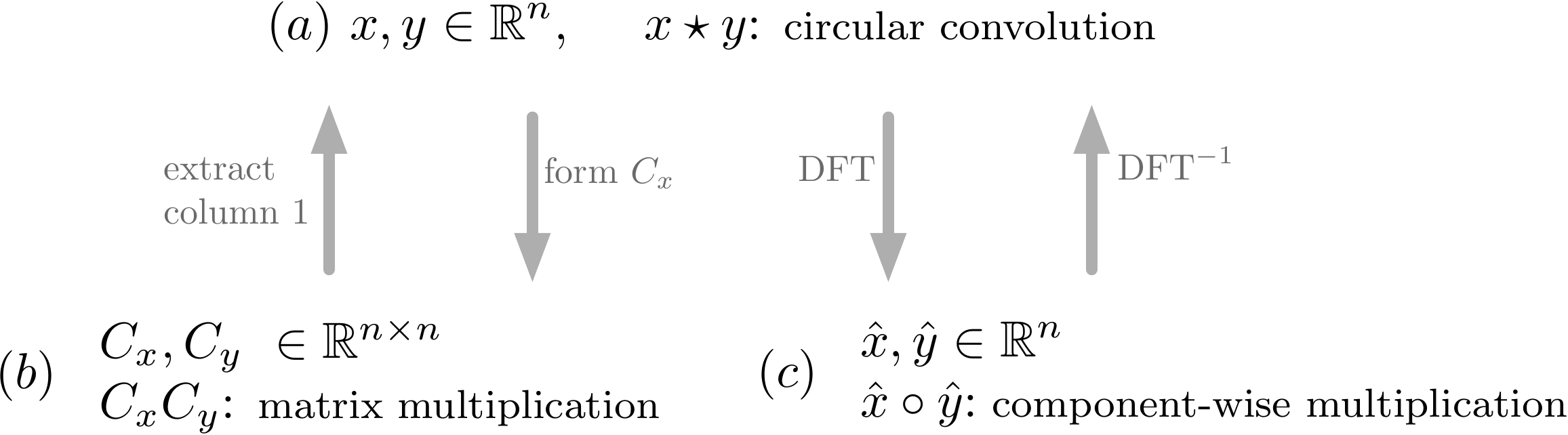}
		\end{center}

%
%
\end{theorem}

\section{Further Comments and Generalizations} 								\label{further.sec}

We end by briefly sketching two different ways in which the procedures described in this tutorial  can be generalized. The first is generalizations to families of mutually commuting infinite matrices and linear operators. These families are characterized by commuting with shifts of functions defined on ``time-axes'' which can be identified with groups or semi-groups. This yields the familiar Fourier transform, Fourier series, and the z-transform.  A second line of generalization is to families of matrices that do not commute. In this case we can no longer demand simultaneous diagonalization, but rather simultaneous {\em block diagonalization} whenever possible. This is the subject of group representations, but we will only touch on the simplest of examples by way of illustration.  
The discussions in this section are meant to be brief sketches to motivate the interested reader into further exploration of the literature.

	\subsection{Fourier Transform,  Fourier Series, and the z-Transform} 						\label{transform.sec}

	First we recap what these classical transforms are. They are summarized in~\cref{trans.tbl}. In a Signals and Systems 
	course~\cite{oppenheim1983signals}, 
	these concepts are usually introduced as transforms on temporal signals, so we will use that language to 
	refer to the independent variable as time, although it can have any other interpretation. As is the theme of this tutorial, 
	the starting point should not be the signal transform, but rather the  systems, or operators, that act on them and their 
	respective invariance properties. We now formalize these properties. 
	\begingroup
		\setlength{\tabcolsep}{4pt} 
	\begin{table}[hhh]
			\begin{center}
			\bgroup									
			\begin{tabular}{|l|l|l|} 	\hline
				\rowcolor{blue!16}
					\btbl \rowcolor{blue!15}	{\bfseries\sffamily	\large Time  \rule{0em}{1em}} 
											\\ {\bfseries\sffamily	\large Axis } \et
							& 
					{\bfseries\sffamily	\large  Transform}  & 
					\btbl  {\bfseries\sffamily	\large  Frequency Axis  } \\
						 {\bfseries\sffamily	  (Frequency ``Set'')  } 	\et
															\\				 \hline \hline 
				\rowcolor{green!2}
				$t\in\R$ & 
					\begin{tabular}{l}	\rule{0em}{1em}
						\textsf{ Fourier Transform}  \\
						$\displaystyle F(\omega) := \int_{-\infty}^{\infty} \!\!\! f(t)~e^{-j\omega t} dt$	
					\end{tabular}
					& \btbl  $j\omega \in j\R$ 	\\ \em imaginary axis of $\C$ \et		\\		 \hline
					\rowcolor{green!2}
					\multicolumn{3}{|c|}{	
							\begin{tabular}{m{0.35\textwidth}m{0.05\textwidth}m{0.35\textwidth}}	
							\includegraphics[height=0.07\textheight]{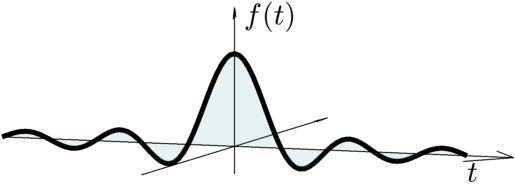}\
							&
							&
							\includegraphics[height=0.07\textheight]{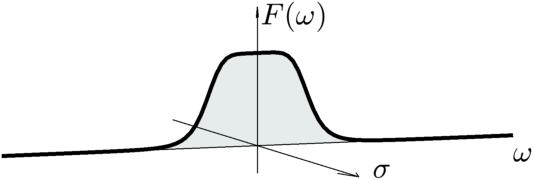}
							\end{tabular}
						}													\\			\hline\hline
				\rowcolor{blue!2}
				$t\in\Z$ & 
					\begin{tabular}{l}	\rule{0em}{1em}
						 \textsf{z-Transform} (bilateral)  \\
						$\displaystyle F(z) := \sum_{t\in\Z} f(t)~z^{- t} 
						$	
					\end{tabular}
					&  \btbl  $z=e^{j\theta}\in\T$ \\ \em unit circle of $\C$ \et 			\\ 	[15pt]	\hline
					\rowcolor{blue!2}
					\multicolumn{3}{|c|}{		
							\begin{tabular}{m{0.35\textwidth}m{0.05\textwidth}m{0.35\textwidth}}	
							\includegraphics[height=0.07\textheight]{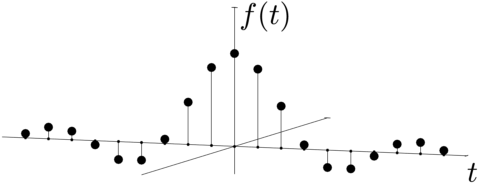}\
							&
							&
							\includegraphics[height=0.07\textheight]{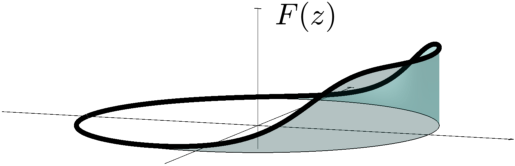}
							\end{tabular}
						}													\\			\hline\hline
				\rowcolor{green!2}
				$t\in\T$ & 
					\begin{tabular}{l}	\rule{0em}{1em}
						\textsf{Fourier Series}  \\
						$\displaystyle F(k) := \int_{0}^{2\pi} \!\!\! f(t)~e^{-jk t } dt$	
					\end{tabular}
					&   \btbl $jk \in j\Z$, ~~~ \em integers of \\ \em imaginary axis of $\C$		\et		\\ 	[18pt]	\hline
					\rowcolor{green!2}
					\multicolumn{3}{|c|}{		
							\begin{tabular}{m{0.35\textwidth}m{0.05\textwidth}m{0.35\textwidth}}	
							\includegraphics[height=0.07\textheight]{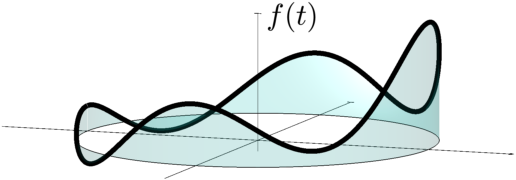}\
							&
							&
							\includegraphics[height=0.07\textheight]{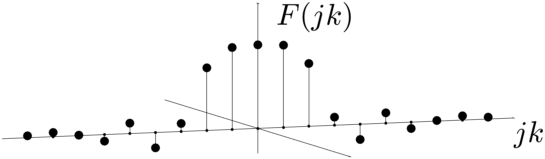}
							\end{tabular}
						}													\\			\hline\hline
				\rowcolor{blue!2}
				$t\in\Z_n$ & 
					\begin{tabular}{l}	\rule{0em}{1em}
						\textsf{Discrete Fourier Transform (DFT)}  \\
						$\displaystyle F(k) := \sum_{t=0}^{n-1} f(t)~e^{-j \frac{2\pi}{n} kt  }$	
					\end{tabular}
					&   \btbl $k \in \Z_n$  \\ \em $n$-roots of unity \et 			\\ 				[18pt]	\hline
					\rowcolor{blue!2}
					\multicolumn{3}{|c|}{		
							\begin{tabular}{m{0.35\textwidth}m{0.05\textwidth}m{0.35\textwidth}}	
							\includegraphics[height=0.07\textheight]{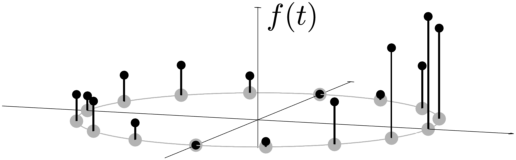}\
							&
							&
							\includegraphics[height=0.07\textheight]{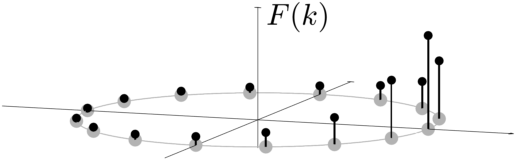}
							\end{tabular}
						}															\\		\hline
			\end{tabular} 
			\egroup
			\end{center}
		\mycaption{A list of Fourier-type transforms. The left column lists the time axis over which a signal 
			is defined, the middle column lists the common name and expression for the transform, and the 
			right column lists the frequency axis, or more precisely the ``set of frequencies'' associated with 
			that transform. The set of frequencies is typically identified with 
			a subset of the complex plane $\C$. Note how all the transforms have the common form of 
			integrating or summing the given signal against  a function of the form $e^{-st}$, 
			where the set of values (``frequencies'') of $s\in\C$ is different for different transforms.   } 
		\label{trans.tbl} 
	\end{table} 
	\endgroup

	The time axes are the integers $\Z$ for discrete time and the reals $\R$ for continuous time. 
	Moreover,  
	the discrete circle $\Z_n$ and the continuous circle $\T$ are the time axes for discrete and continuous-time periodic 
	signals respectively. 
	A common feature of  the time axes $\Z$, $\R$, $\T$ and $\Z_n$ is that they all are {\em commutative groups}. In fact, 
	they are the basic commutative groups. All other (so-called locally compact) commutative groups are made up of 
	group products of those basic four~\cite{rudin2017fourier}.
	
	Let's denote by $t,\tau,T$  elements of those groups $\G = \R, ~\Z, ~\T, \mbox{ or } \Z_n$. For any function $f:\G\rightarrow \R$
	(or $\C$) defined on such a group, there is a natural ``time shift'' operation 
	\be
		\left(S_T f \right)(t) ~:=~ f(t-T), 
	  \label{t_shift.eq}	
	\ee
	which is the right shift (delay) of $f$ by $T$ time units. All that is needed to make sense of this operation is that for 
	$t,T\in\G$, we have $t-T\in\G$, and that is guaranteed by the group structure for any of those  four time sets. Now 
	consider a linear operator $A:\C^\G \longrightarrow \C^\G$ acting on the vector space $\C^\G$ of all scalar-valued 
	functions on $\G$, which can be any of the four time sets. We call such an operator {\em time invariant} (or {\em shift invariant})
	if it commutes with all possible  time-shift operations~\req{t_shift}, i.e. 
	\be
		\forall T\in\G, ~~~~ S_T A ~=~ A S_T. 
	  \label{shift_inv.eq}
	\ee
	To conform with traditional terminology, we refer to such shift-invariant linear operators as {\em Linear Time-Invariant (LTI)
	systems}. They are normally described as differential or difference 
	equations with a forcing term, or as convolutions of signals amongst 
	other representations. 
	However, only the shift-invariance property~\req{shift_inv}, and not the details of those representations, is what's important 
	in discovering the appropriate transform that simultaneously diagonalizes such operators. 
	
	There are additional technicalities in generalizing the previous techniques to 
	sets of {\em linear operators}
	on infinite dimensional spaces rather 
	than matrices. The procedure however is very similar. We identify the class of operators to be analyzed. This involves
	a shift (time) invariance property, which then implies that they all mutually commute. The ``eigenvectors'' of the shift operators
	  give the
	simultaneously diagonalizing  transform. The complication here is that eigenvectors may not 
	exists in the classical sense (they do in the case of Fourier series, but not in the other cases). In addition, diagonalization 
	will not necessarily correspond to finding a new basis of the vector space. In both the Fourier and z-transforms, the 
	number of ``linearly independent eigenfunctions'' is not even countable, so they can't be thought of as forming a basis. 
	Fortunately, it is  easy to 
	circumvent these difficulties by generalizing the concept 
	of diagonal matrices to {\em multiplication operators}. For linear operators on infinite-dimensional spaces, these 
	play the same role as the  diagonal matrices do on finite-dimensional spaces.

		\begin{definition} 
			Let $\Omega$ be some set, and consider the vector space $\C^\Omega$ 
			of all scalar-valued functions on $\Omega$. 
			Given a particular scalar-valued function $a:\Omega \longrightarrow \C$, we define the associated 
			{\em multiplication operator} $M_a: \C^\Omega \longrightarrow \C^\Omega$ by 
			\[
				\forall x\in\Omega, ~~~~\lb M_a f \rb(x) ~:=~ a(x) ~f(x) ,  
			\]
			i.e. the point-wise multiplication of $f$ by $a$. 
		\end{definition} 
		If $\Omega=\lcb 1, \ldots, n \rcb$, then $\C^\Omega= \C^n$, the space of all complex $n$-vectors, and $M_a$
		is simply represented by the diagonal matrix whose entries are made up of the entries of the vector $a$. 
		The concept introduced above is however much more general. 
		Note that it is immediate from the definition that all multiplication operators mutually commute, just like 
		all diagonal matrices mutually commute. 
		
		Now we generalize the concept of diagonalizing matrices. Diagonalizing an operator, when possible, is done 
		by converting it to a multiplication operator. 
		\begin{definition} 														\label{diag.def}
			A linear operator $A:\cH  \longrightarrow \cH $ on a vector space $\cH$ is said to be {\em diagonalizable} if 
			there exists a function space $\C^\Omega$, and an invertible 
			transformation $V:\C^\Omega \longrightarrow \cH$ that 
			converts $A$ into a multiplication operator $M_a$ 
			\[
				V A V^{-1} ~=~ M_a, 
			\]
			for some function $a:\Omega\longrightarrow \C$. The function $a$ is referred to as the {\em symbol} 
			of the operator $A$, and $V$ is the {\em diagonalizing} transformation. 
		\end{definition} 
		Thus in contrast to the case of matrices, we may have to move to a different vector space to diagonalize a 
		general linear operator. 
		
		In some cases, we can still give a diagonalization an interpretation as a basis expansion in the same 
		vector space. It provides  a helpful contrast to consider such an example. 		
		Let an operator  $A:\cH\rightarrow\cH$ have a 
		  countable set of eigenfunctions 
		$\left\{ v_m\right\}$ that span a Hilbert space $\cH$. Assume in addition that $A$ is normal, 
		and thus the eigenfunctions are mutually orthonormal. An example of this situation 
		 the case of shift invariant operators on $\T$, which corresponds to the 3rd entry in~\cref{trans.tbl} 
		(i.e. Fourier series). In that case we can take  $\Omega=\Z$, and thus $\C^\Z$ is the space of all complex-valued
		bilateral sequences. In addition we can add a Hilbert space structure by using $\ell^2$ norms and consider $\ell^2(\Z)$ 
		as the space of sequences. 
		The diagonalizing 
		transformation\footnote{In this case the function space is $\C^\N$, the space of semi-infinite sequences. 
			We need to add a Hilbert space structure in order to make sense of converges of infinite sums, 
			and the choice $\ell^2(\N)$ provides additional nice properties such as Parseval's identity. We do 
			not discuss these here. 
				} 
		is $V: \ell^2(\Z) \longrightarrow L^2(\T)$ 
		described as follows. 
		Let $v_k(t) := e^{jkt}/\sqrt{2\pi}$ be the Fourier series elements. They are an orthonormal basis of $L^2(\T)$. 
		Consider  any square-integrable  function $f\in L^2(\T)$ with Fourier series 
		\[
			F_k := \inprod{v_k}{f} = \frac{1}{\sqrt{2\pi}} \int_0^{2\pi}  \!\!\! e^{-jkt} ~f(t)~dt ,
			~~~~~~
			f(t) = \sum_{k\in\Z} F_k ~v_k(t)   = \frac{1}{\sqrt{2\pi}} \sum_{k\in\Z} F_k e^{jkt} .
		\]
		The mapping $V: \ell^2(\Z) \longrightarrow L^2(\T)$ then simply maps each function $f$ to its bilateral Fourier series
		coefficients 
		\[
			\left\{\ldots, F_{\sm 1},   F_0, F_1 , \ldots \right\} 
			~~~\leftrightarrow~~~ 
			f = \sum_{k\in\Z} F_k  v_k. 
		\]
		Plancherel's theorem guarantees that this mapping is a bijective isometry. Any shift-invariant operator $A$ on $L^2(\T)$ 
		(e.g. those that can be written as circular convolutions) 
		then has a diagonalization as a doubly infinite matrix 
		\[
			VAV^{-1} ~=~ \left[ \begin{smallmatrix} 	\ddots 	& 			& 		&  			&		\\
									&	\ah_{-1} 	& 		& 			&		\\ 
									&			& \ah_{0}	&			&		\\
									&			&		& \ah_{\sm 1}	&		\\	
									&			&		&			& \ddots
						\end{smallmatrix}  	\right],
		\]
		where the sequence $\left\{ \ah_{k} \right\}$ is made up of the eigenvalues of $A$.

		The example (Fourier series) just discussed is a  very special case. 
		In general, we have to consider diagonalizing using a
		multiplication operator on an uncountable domain. Examples of these are the Fourier and z-transforms, where the
		diagonalizations are multiplication operators on $\C^\R$ and $\C^\T$ respectively. Note that both sets $\R$ and $\T$ 
		are uncountable, thus an interpretation of the transform as a basis expansion is not possible. None the less, 
		multiplication operators provide the necessary generalization of diagonal matrices.

\subsection{Simultaneous Block-Diagonalization and Group Representations} 							\label{s3.sec}

	The simplest example of a non-commutative group of transformations is the so-called {\em symmetric group} $\mbS_3$ of 
	all permutations of ordered $3$-tuples. Consider the ordered $3$-tuple $\lb 0,1,2\rb$ and the following ``circular shift'' and
	``swap'' operations on it 
\be
		\begin{array}{rcl} 
            		I  \lb 0,1,2\rb &:=& \lb 0,1,2 \rb,		\\
            		c \lb 0,1,2\rb &:=& \lb 2,0,1 \rb, 		\\ 
            		c^2 \lb 0,1,2\rb &:=& \lb 1,2,0 \rb, 
		\end{array}
		~~~~~~~
		\begin{array}{rcl} 
            		s_{01} \lb 0,1,2 \rb &:=& \lb 1, 0 , 2\rb, 	\\
            		s_{12} \lb 0,1,2 \rb &:=& \lb 0, 2 , 1\rb, 	\\
            		s_{20} \lb 0,1,2 \rb &:=& \lb 2, 1 , 0\rb, 
		\end{array}
  \label{perm.eq}
\ee
where $I$ is the identity (no permutation) operation, and $s_{ij}$ is the operation of swapping the $i$ and $j$ elements.  The group operation is the composition of permutations. Note that the 
first three permutations $\left\{ I, c, c^2 \right\}$ are isomorphic to $\Z_2$ as a group and thus mutually commute. The swap and 
shift operations in general do not mutually commute. A little investigation shows that the six elements 
\[
	\left \{ I, ~c , ~c^2, ~ s_{01}, ~s_{12}, ~ s_{20} \right\} ~=~ \mbS_3 
\]
do indeed form the group of all permutations of a $3$-tuple. 

A {\em  representation}  of a group is an isomorphism between the group and a set of matrices (or linear operators) with the 
composition operation between them being 
standard matrix multiplication. With a slight abuse of notation (where we use the same symbol for the group element and its
representer) we have the following representation of $\mbS_3$ as linear operators on $\R^3$ (i.e. as matrices on $3$-vectors) 
\be
                	I = \bsm	1 & 0 & 0 	\\ 
                				0 & 1 & 0 \\ 
                				0 & 0 & 1 \esm,
                	c = \bsm 		0 & 1 & 0 	\\ 
                				0 & 0 & 1 \\ 
                				1 & 0 & 0 \esm, 
                	c^2 = \bsm 0 & 0 & 1 	\\ 
                				1 & 0 & 0 \\ 
                				0 & 1 & 0 \esm,			\\
                	s_{01} = \bsm 0 & 1 & 0 	\\ 
                				1 & 0 & 0 \\ 
                				0 & 0 & 1 \esm,
                	s_{12} = \bsm 1 & 0 & 0 	\\ 
                				0 & 0 & 1 \\ 
                				0 & 1 & 0 \esm,
                	s_{20} = \bsm 0 & 0 & 1 	\\ 
                				0 & 1 & 0 \\ 
                				1 & 0 & 0 \esm		.
  \label{s3.eq}
\ee
Those matrices acting on a vector $\lb x_0,x_1,x_2\rb$ will permute the elements of that vector according to~\req{perm}. 

Now we ask the question: {\em  can the set of matrices in $\mbS_3$ (identified as~\req{s3}) be simultaneously diagonalized?} 
Recall that commutativity 
is a necessary condition for simultaneous diagonalizability, and since this set is not commutative, the answer is no.
 The 
failure of commutativity can be  seen from the following easily established relation between shifts and 
swaps 
\[
	s_{kl} c ~=~ c ~ s_{k+1,l+1}, ~~~~~~k,l\in\Z_2
\]
(i.e. the arithmetic for $k+1$ and $l+1$ should be done in $\Z_2$). 

The lack of commutativity precludes simultaneous 
diagonalizability. However, it is possible to simultaneously block-diagonalize all elements of $\mbS_3$ so they all have the following 
block-diagonal form
\be
	\otbm{*}{0}{0}{0}{*}{*}{0}{*}{*} . 
  \label{bdform.eq}
\ee
In some sense, this the simplest form one can hope for when analyzing all members of $\mbS_3$ (and the algebra generated by it). 
This block diagonalization does indeed reduce the complexity of analyzing a set of $3\times 3$ matrices to analyzing sets of at most 
$2\times 2 $ matrices. While this might not seem significant at first, it can be immensely useful in certain cases. Imagine for example 
doing symbolic calculations with $3\times 3$ matrices. This typically yields unwieldy formulas. A reduction to symbolic calculations 
for $2\times 2$ matrices can give significant simplifications. Another case is when infinite-dimensional operators can be block
diagonalized with finite-dimensional blocks. This is the case when one uses Spherical Harmonics to represent rotationally invariant 
differential operators. In that case the representation has finite-dimensional blocks, though with increasing size. 

\subsubsection*{Block Diagonalization and Invariant Subspaces} 

Let's first examine how block diagonalization can be interpreted geometrically. Given an operator $A:\cV \longrightarrow \cV$ on a vector space, we say that a subspace $\cV_o\subseteq \cV$ is {\em $A$-invariant} if $A\cV_o\subseteq\cV_o$ (i.e. for any $v\in\cV_o$, $Av\in\cV_o$). Note that the span of any eigenvector of $A$ (the so-called eigenspace) is an invariant subspace of dimension 1.   Finding invariant subspaces is equivalent to {\em block triangularization}. Let $\cV_1$ be some complement of $\cV_o$ (i.e. $\cV = \cV_o \oplus \cV_1$), then with respect to that decomposition, $A$ has the form 
\be
	A ~=~ \tbtdash{A_{11}}{A_{12}}{0}{A_{22}} . 
   \label{A_triang.eq}
\ee
Note that in general, the complement subspace will not be $A$-invariant. If it were,  then $A_{12}=0$ above, and that form of $A$ would be block diagonal. Thus {\em block diagonalization amounts to finding an $A$-invariant subspace $\cV_o$, as well as a complement $\cV_1$ of it such that $\cV_1$ is also $A$-invariant}. 

Now observe the following facts which are immediately obvious (at least for matrices) 
from the the form~\req{A_triang}. If $A$ is invertible, then $\cV_o$ is also $A^{-1}$-invariant since the inverse of an upper-block-triangular matrix is also upper-block-triangular. If we choose $\cV_1=\cV_o^\perp$, the orthogonal complement of $\cV_o$, then $\cV_o$ is $A$-invariant iff $\cV_o^\perp$ is $A^*$-invariant (this can be seen from~\req{A_triang} by observing that $A^*$ is block-lower-triangular). Finally, in the special 
case that  $A$ is unitary (i.e. $AA^*=A^*A=I$, and therefore $A^{-1}=A^*$),  it follows from the previous two observations that for a unitary $A$, any $A$-invariant subspace $\cV_o$ is such that its orthogonal complement $\cV_o^\perp$ is automatically $A$-invariant. Therefore, for unitary matrices, block triangularization is equivalent to block diagonalization, which can be done by finding invariant subspaces and their orthogonal complements. 

\subsubsection*{Block Diagonalization of $\mbS_3$} 

Now we return to the matrices of $\mbS_3$~\req{bdform} and show how they can be simultaneously block diagonalized. Note that all the matrices are unitary, and therefore once  all the common invariant subspaces are found, they are guaranteed to be mutually orthogonal. 
 The easiest one 
to find is the vector $\lb 1, 1, 1\rb$. Note that it is an eigenvector of all members of $\mbS_3$ with eigenvalue $1$ (since obviously any permutation of the elements of this vector produce the same vector again). This is an 
eigenspace of dimension $1$. There is not another shared eigenspace of dimension $1$ since then we would have simultaneous 
diagonalizability, and we know that is precluded by the lack of commutativity. We thus have to simply find the 2-dimensional
 orthogonal complement  
of the span of $\lb 1, 1, 1\rb$. There are several choices for its basis. One of them is as follows 
\[
	v_1 := \bbm 1 \\ 1 \\ 1 \ebm , 
	~~~
	v_2 := \bbm \sm 1 \\ 0 \\ 1 \ebm , 
	~~~
	v_3 := \bbm 0 \\ \sm1 \\ 1 \ebm. 
\]
Notice that the vectors $\{v_i\}$ are mutually orthogonal, which simplifies calculations that finally give the elements of $\mbS_3$ 
in this new basis as 
\be
                	c = \otbms{1}{0}{0}
                				{0}{\sm1}{\sm1}
                				{0}{1}{0}, 
                	c^2 = \otbms{1}{0}{0}
                				{0}{0}{\sm1}
                				{0}{\sm1}{\sm1}, 
                	s_{01} =\otbms{1}{0}{0}
                				{0}{0}{1}
                				{0}{1}{0},
                	s_{12} = \otbms{1}{0}{0}
                				{0}{1}{\sm 1}
                				{0}{1}{ \sm 1},
                	s_{20} = \otbms{1}{0}{0}
                				{0}{\sm1}{ \sm 1}
                				{0}{0}{1},
  \label{s3_bd.eq}
\ee
which are all indeed of the form~\req{bdform}. 

It is more common in the literature to perform the above analysis in the language of group representations, specifically as decomposing a given representation into its component {\em irreducible representations}. Block diagonalization is then an observation about the matrix form that the representation takes after that decomposition. For the student proficient in linear
algebra, but perhaps not as familiar with group theory, a more natural motivation is to start as done above from the block-diagonalization problem as the goal, and then use group representations as a tool to arrive at that goal. 

What has been done above can be restated using group representations as follows. A representation of a group $\G$ is a 
group homomorphism   $\rho:\G \longrightarrow \GL(V)$ into the group $\GL(V)$ of invertible linear transformations of a vector space $V$. Assume for simplicity that $\G$ is finite, $\rho$ is injective,  $V$ is finite dimensional, and that all transformations $\rho(\G)$ are unitary.
The matrices~\req{s3} of $\mbS_3$ are in fact the images of an injective, unitary homomorphism $\rho:\mbS_3 \longrightarrow \GL(3)$ into the group of all non-singular transformations of $\R^3$.  

A representation is said to be {\em irreducible} if there are no non-trivial  invariant subspaces common to all transformations $\rho(\G)$. In other words, all elements of $\rho(\G)$ cannot be simultaneously 
block diagonalized. As we demonstrated,~\req{s3} is indeed reducible. More formally, let $\rho_i:\G\longrightarrow \GL(V_i)$, $i=1,2$ be two given representations. Their {\em direct sum} $\rho_1\oplus\rho_2: \G \longrightarrow \GL(V_1\oplus V_2)$ is the representation formed by the ``block diagonal'' operator 
\be
	\bbm \rho_1(g) & 0 \\ 0 & \rho_2(g) \ebm , 
  \label{two_rep.eq}
\ee
with the obvious generalization to more than two representations. If a representation is reducible, then the existence of a common invariant subspace means that it can be written as the direct sum of so-called ``subrepresentations'' as in~\req{two_rep}. Thus 
{\em simultaneous block-diagonalization into the smallest dimension blocks} is equivalent to the {\em decomposition of a given representation into the direct sum of irreducible representations}. This is what we have done for the representation~\req{s3} of $\mbS_3$ by finding the two common invariant subspaces (which contain no proper   further subspaces that are invariant) and thus brining all of them into the block diagonal form~\req{s3_bd}.  In general, it is a fact that any representation of a finite group (more generally, of a compact group) can be decomposed as the direct sum of irreducible representations~\cite{diaconis1988group,serre2012linear}.

\bibliographystyle{IEEEtran} 
\bibliography{refs}

\appendix

\section{Exercises}


	\subsection{Circulant Structure} 
												\label{circ_shift.ex}
	Show that any matrix $M$ that commutes with the shift operator $S$ must be a circulant matrix, i.e. must have the structure shown in~\req{circ_def}, or equivalently~\req{circ_def_ind}.  

	\textbf{Answer:} 
			Starting from the relation $SM=MS$, and using the definition  $(S)_{ij} = \delta_{i-j-1}$ compute 
                    \begin{align*}
                    	\lb  S M \rb_{ij}	& = 
				\sum_l S_{il} \lb M \rb_{lj} = \sum_l \delta_{i-l-1} ~ \lb M \rb_{lj}  =  \lb M \rb_{i-1,j}  ,	\\
                    	\lb   M S \rb_{ij}	& = 
				\sum_l \lb M \rb_{il}  S_{lj}  = \sum_l \lb M \rb_{il}  ~\delta_{l-j-1} = \lb M \rb_{i,j+1}   . 
                    \end{align*}
		Note that since the indices $i-j-1$ of the Kroenecker delta are to be interpreted using modular arithmetic, then the 
		indices $i-1$ and $j+1$ of $M$ above should also be interpreted with modular arithmetic. The statements
		\[
			\lb M \rb_{i-1,j} = \lb M \rb_{i,j+1}
			~~\Leftrightarrow~~
			\lb M \rb_{i-1,j-1} =\lb M \rb_{i,j}
			~~\Leftrightarrow~~
			\lb M \rb_{i,j} =\lb M \rb_{i+1,j+1}
		\]
		then mean that the $i$'th column is obtained from the previous $i-1$ column by circular right shift of it. 
		
		Alternatively, 
		the last statement above implies that for any $k$, $\lb M \rb_{i,j} =\lb M \rb_{i+k,j+k}$, i.e. that entries of $M$ are constant 
		along ``diagonals''. Now take the first column of $M$ as 
		$m_i := \lb M \rb_{i,0}$, then 
		\[
			\lb M \rb_{ij} ~=~ \lb M \rb_{i-j,j-j} ~=~ \lb M \rb_{i-j,0} ~=~ m_{i-j} .
		\]	
		Thus all entries of $M$ are obtained from the first column by circular shifts as in~\req{circ_def_ind}.


	\subsection{co-Prime Powers of the Shift}									
															\label{p_prime.ex}
	Show that an $n\times n$ matrix $M$ is circulant iff it commutes with $S^p$ where $(p,n)$ are coprime.   

	\textbf{Answer:} 
		If $M$ is circulant, then it commutes with $S$ and also commutes with any of its powers $S^p$. The other direction is
		more interesting. 
		
		The basic underlying fact for this conclusion has to do with modular arithmetic in $Z_n$. If $(p,n)$ are coprime, then there 
		are integers $a$, $b$ that satisfy the Bezout identity 
		\[
			ap+bn ~=~1, 
		\] 
		which also implies that $ap$ is equivalent to $1$ mod $n$ since $ap = 1-bn$, i.e. it is equal to a multiple of $n$ plus $1$. 
		Therefore, there exists a power of $S^p$, namely $S^{ap}$ such that 
		\be
			S^{ap} ~=~ S. 
		  \label{Sap.eq}
		\ee
		Thus if $M$ commutes with $S^p$, then it commutes with all of its powers, and namely with $S^{ap}=S$, i.e. it commutes
		with $S$, which is the condition for $M$ being circulant. 
		
		Equation~\req{Sap} has a nice geometric interpretation. $S^p$ is a rotation of the circle in~\cref{shift_inv.fig} by $p$ 
		steps. If $p$ and $n$ were not coprime, then regardless of how many times the rotation $S^p$ is repeated, there will be some 
		elements of the discrete circle that are not reachable from the $0$ element by these rotations 
		(examine also~\cref{eigs_viz.fig} for an illustration of this). 
		The condition $p$ and $n$ coprime insures that there
		is some repetition of the  rotation $S^p$, namely $\lb S^{p}\rb^a$ which gives the basic rotation $S$. Repetitions of $S$ then of 
		course generate all possible rotations on the discrete circle. In other words, $p$ and $n$ coprime insures that by repeating 
		the rotation $S^p$, all elements of the discrete circle are eventually reachable from $0$.


	\subsection{Commutativity and Associativity} 							
																\label{conv_ass_com.ex}
	Show that circular convolution~\req{circ_conv} is  commutative and associative. 

	\textbf{Answer:} 	
		\noindent
		\textit{Commutativity:}  Follows from 
		\[
			\lb a \star b \rb_k ~=~ \sum_{l} a_l b_{k-l} ~=~ 
				 \sum_{l} a_{k-j} b_{j} ~=~ \lb b \star a \rb_k 
		\]
		where we used the substitution $j=k-l$ (and consequently $l=k-j$) . 
		
		\noindent
		\textit{Associativity:} First note that $\lb b \star c \rb_i = \sum_j b_j c_{i-j} $, and compare 
			\begin{align*}
				\lb a \star \lb b \star c \rb \rb_{k} &=~ \sum_{l} a_l \lb b\star c  \rb_{k-l}
					~=~  \sum_{l} a_l \lb \sum_{j} b_{j} c_{k-l-j}    \rb			
					~=~  \sum_{l,j} a_l b_{j} c_{k-l-j}    			\\
				\lb \lb a \star  b \rb \star c  \rb_{k} &=~ \sum_{j} \lb a \star b \rb_j  c_{k-j} 
					~=~  \sum_{j}  \lb   \sum_{l} a_l  b_{j-l} \rb   c_{k-j}    	
					~=~  \sum_{l,j}    a_l  b_{j-l}   c_{k-j}    	.
			\end{align*} 
		Relabeling $j-l =: i$ (and therefore  $j=l+i$) in the second sum makes it 
		\[
				\sum_{l,i}    a_j  b_{i}   c_{k-(l+i)}    
				~=~  \sum_{l,i}    a_j  b_{i}   c_{k-l-i}    , 
		\]
		Which is exactly the first sum, but with a different labeling of the indices. 

\end{document}